\documentclass[12pt,a4paper]{article}
\usepackage[english]{babel}
\usepackage[centertags]{amsmath}
\usepackage{amsfonts}
\usepackage{amssymb}
\usepackage{amsthm}
\usepackage{amscd}
\newcommand{\dis}{\displaystyle}
\newcommand{\ra}{\rangle}
\newcommand{\la}{\langle}

\newcommand{\opAv}{\hat{\mathbf{A}}}

\newcommand{\opB}{\hat{{B}}}
\newcommand{\opC}{\hat{{C}}}
\newcommand{\opa}{\hat{a}}
\newcommand{\opb}{\hat{b}}
\newcommand{\opc}{\hat{c}}
\newcommand{\opd}{\hat{d}}

\newcommand{\bd}{\mathbf{d}}

\newcommand{\bk}{\mathbf{k}}

\newcommand{\bn}{\mathbf{n}}

\newcommand{\bS}{\mathbf{S}}

\newcommand{\bu}{\mathbf{u}}
\newcommand{\br}{\mathbf{r}}

\newcommand{\bepsilon}{\boldsymbol{\epsilon}}

\newcommand{\ing}{{\text{In}}}
\newcommand{\out}{{\text{Out}}}
\newcommand{\el}{\mathcal{L}}
\newcommand{\er}{\mathcal{R}}
\newcommand{\bes}{\begin{split}}
\newcommand{\ens}{\end{split}}

\begin{document}

\parindent=20pt

\begin{center}{\Huge

Input-Output Relations in Optical  \\ [5mm] Cavities: a Simple
Point of View }

\vfill

ANDREA AIELLO\footnote{e-mail address:
andrea.aiello@roma1.infn.it} \\ [5mm]

Dipartimento di Fisica, Universit\'{a} degli Studi di Roma \\
[3mm] {\em "La Sapienza"}, P.le A. Moro 2, 00185 Rome, Italy. \\
[5mm]

\vfill
\end{center}

\baselineskip=16pt
%

\thispagestyle{empty}

\newpage
\pagestyle{empty} \centerline{ \huge \bf Abstract}
 In this work we present a very simple approach to input-output
 relations in optical cavities, limiting ourselves to one- and two-photon
 states of the field.
 After field quantization, we derive
  the non-unitary transformation between {\em Inside} and {\em Outside}
 annihilation and creation operators. Then we express the most general
 two-photon state generated by {\em Inside} creation operators,
through base states generated by {\em Outside} creation operators.
  After renormalization
 of coefficients of inside two-photon state,
 we calculate the outside photon-number probability distribution
 in a general case. Finally we treat with some detail the single
 mode and symmetrical cavity case.
\bigskip
\bigskip
\bigskip
\bigskip
\bigskip


\newpage
\pagestyle{plain} \setcounter{page}{1}
%
%
\section{Introduction}
The problem of interaction between a pair of atoms in free space
and in a cavity has been subject of several investigations in the
past years \cite{Fermi,Milonni,Biswas,Aiello,fdm1,fdm2}. Recently
the spontaneous emission of a pair of two identical atoms or
molecules in a planar Fabry-P\'erot microcavity has been subject
of theoretical and experimental research
\cite{Takada,deangelis99}. This work starts from our attempt to
give
 a simple interpretation to some recent experimental results.
Consider  a process of spontaneous emission of a pair of photons
by two distinct molecules inside a microcavity. If we measure the
number of photons emitted outside the cavity, and if there are not
dissipative phenomena, we will find only three possible results:
two photons detected on the right and no one on the left, two
photons detected on the left and no one on the right, one photon
detected on the left and one on the right. In this work we look
for a solution to this question: if we know the distribution of
the number of photons outside the cavity, how can we obtain
information about the process of emission which generated the
photons without completely solving the problem? Our basic idea is
to describe the electromagnetic field inside the cavity by means
of a two-photon state as general as possible; the coefficients of
the expansion of this state in a proper basis set  depend on the
process which generated the state itself.
 If we project this state on the basis number-states
defined outside the cavity, we obtain directly the probability
distribution we look for. Therefore if we change the above
mentioned coefficients, we can study how the probability
distribution changes, and by comparison with
 the measured distribution we can obtain the values of these coefficients
getting information about the process active inside the cavity.
This procedure is not orthodox or fully justified, however the
obtained results  are in qualitative agreement with experimental
results.
 \\ The relations of Input-Output and their connection with the traditional
  stochastic methods
based on Langevin equations have been studied by Kn\"{o}ll {\em et
al.} \cite{knoll}. The aim of the present paper is to generalize
the methods used to describe an optical element with two inputs
and two outputs, as a Beam-Splitter, to derive simple relations
between fields inside and outside a planar Fabry-P\'erot cavity.
In Sec. II we quantize the electromagnetic field generalizing to
three dimensions an approach presented by Barnett {\em et al.}
\cite{barnett96} for one-dimensional fields. In Sec. III we
restrict our attention to one-dimensional fields and derive the
relations between operators defined in the space Inside and
Outside the cavity. Then, in Sec. IV we build and study the states
of the field generated by linear and bilinear forms of creation
operators both inside and outside the cavity and, in Sec. V, we
calculate the outside probability distributions for these
two-photon states. Finally, we summarize our results in Sec. VI.
\\
\section{Spatial modes and Field Quantization}
Now we calculate the appropriate normal modes for quantization of
the electromagnetic field. In a traditional approach, one first
determines the modes of the classical boundary value problem, then
one quantizes the field in terms of these modes
\cite{ley87,FDMI91}. An alternative approach has been presented by
Barnett {\em et al.} \cite{barnett96}. We have generalized this
work, restricted to an one-dimensional fields, to
three-dimensional fields. \\ In free space, having fixed the
Coulomb gauge, the spatial variations of the electromagnetic field
can be described by the solutions $\bu_k(\br)$ of the Helmholtz
equations
\\
\begin{equation}\label{10}
 \nabla^2 \bu_k ( \br) + k^2 \bu_k (
  \br)= 0 .
\end{equation}
\\
For each value of $k$, there are two solutions that we indicate
with $\bu_{ R \lambda}(\bk,\br)$ and $\bu_{ L \lambda}(\bk,\br)$.
We choose them as plane waves that are incident respectively from
left and right on the plane of equation $z = \mathrm{const.}$, as
shown in Fig. 1. \\ Here $\lambda=1,2$ is the polarization index
and $\bk$ is the wave-vector such as $|\bk|^2 = k^2 \equiv
\omega^2/c^2$. The explicit form of $\bu_{ R \lambda}(\bk,\br)$
and $\bu_{ L \lambda}(\bk,\br)$ is therefore
\\
\begin{equation}\label{20}
 \begin{split}
\bu_{ R \lambda} (\bk, \, \br) & = \bepsilon_\lambda (\bk_+) \exp
(i \bk_+ \cdot \br), \\\\ \bu_{ L \lambda} (\bk, \, \br) & =
\bepsilon_\lambda (\bk_-) \exp (i \bk_- \cdot \br),
 \end{split}
\end{equation}
\\
where we have defined
\\
\begin{equation}\label{30}
 \mathbf{k}_\pm = k (\sin \theta \, \cos \phi,
 \; \sin \theta \, \sin \phi, \; \pm \cos \theta),
\end{equation}
\\
for $(0 \leq \theta \leq \pi /2)$, and
\begin{equation}\label{40}
\begin{split}
\bepsilon_1 (\bk_\pm) & = (\sin \phi, -\cos \phi, 0),
\\\\
 \bepsilon_2 (\bk_\pm) & = (\cos \theta
\cos \phi, \cos \theta \sin \phi, \mp  \sin \theta).
\end{split}
\end{equation}
\\
\\
\begin{picture}(460,240)(-30,30)
\thinlines { \thicklines

 \put(190,150){\oval(40,40)[b]}
 \put(190,50){\line(0,1){220}}
 }
 \put(120,100){\makebox(0,0){$\bk_+$}}
 \put(260,100){\makebox(0,0){$\bk_-$}}
 \put(70,125){\makebox(0,0){$\bu_\lambda(\bk,\br)$}}
 \put(310,125){\makebox(0,0){$\bu_\lambda'(\bk,\br)$}}
 \put(165,125){\makebox(0,0){$\theta$}}
 \put(215,125){\makebox(0,0){$\theta$}}
 \put(175,175){\makebox(0,0){$0$}}
 \put(370,160){\makebox(0,0){$z$}}
 \put(20,160){\vector(1,0){340}}
 \put(70,100){\vector(2,1){120}}
 \put(310,100){\vector(-2,1){120}}
\end{picture}

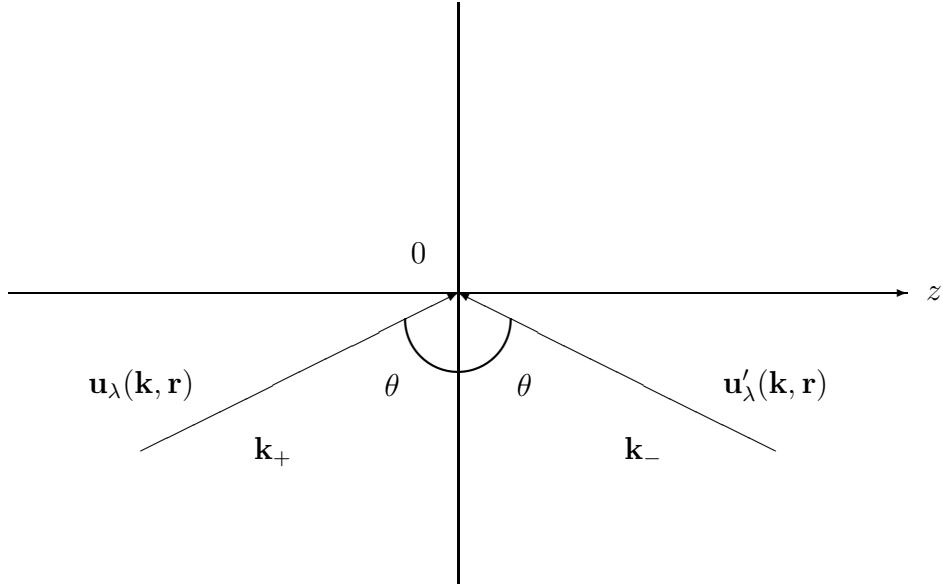
FIG.1.  Scheme of the plane waves $\bu_\lambda(\bk,\br)$ and
$\bu_\lambda'(\bk,\br)$ incident respectively  from left and from
right on the plane $z=0$.
\\
\\
Now, following a standard procedure \cite{loudon,khosravi91}, we
introduce the mode creation and destruction operators $\opa_{R
\lambda}(\bk)$ and $\opa_{L \lambda}(\bk)$ as
\\
\begin{equation}\label{50}
\begin{array}{ccccc}
  \opa_{ R\lambda}^\dag(\bk) & \text{and} & \opa_{ R\lambda} (\bk)
  & \text{for the mode} &  \bu_{ R \lambda}(\bk,\br) ,\\\\
  \opa_{ L\lambda}^{ \dag } (\bk) & \text{and} & \opa_{ L\lambda}(\bk)
  & \text{for the mode}
  &  \bu_{ L \lambda}(\bk,\br).
\end{array}
\end{equation}
\\
Being $\bk$ a three-dimensional continuous variable, these
operators satisfy the following commutation relations:
\\
\begin{equation}\label{60}
 \begin{array}{ccl}
  \left[ \opa_{ R\lambda} (\bk), \,\opa_{ R \lambda'}^\dag (\bk')  \right]  =
  &
  \left[ \opa_{ L\lambda} (\bk), \,\opa_{ L\lambda'}^{  \dag} (\bk')  \right]
  = &  \delta_{\lambda {\lambda'}} \delta (\bk - \bk'), \\\\
 \left[ \opa_{ L\lambda} (\bk), \,\opa_{ R\lambda'}^{  \dag} (\bk')
 \right]= &
 \left[ \opa_{ R\lambda} (\bk), \,\opa_{ L\lambda'}^{  \dag} (\bk')  \right]
   = & 0.
  \end{array}
\end{equation}
\\
The vector potential operator is written, in Heisenberg
representation, as
\\
\begin{equation}\label{70}
 \opAv (\br,t) = \opAv^+ (\br,t) + \opAv^- (\br,t),
\end{equation}
\\
where
\\
\begin{equation}\label{80}
\begin{split}
 \opAv^+ (\br,t)   =  & \int d \bk \left\{ \left( \frac{\hbar
}{16 \pi^3 \varepsilon_0  \omega } \right)^{1/2}  \right. \\\\ &
\times \left. \sum_{\lambda = 1,2} \Bigl[ \bu_{R \lambda} (\bk,\,
\br) \opa_{R \lambda} (\bk) + \bu_{L \lambda} (\bk,\, \br) \opa_{L
\lambda} (\bk) \Bigr]   \exp (- i \omega t) \right\},
\end{split}
\end{equation}
\\
and $\opAv^- (\br,t) = [ \opAv^+ (\br,t)]^\dag $. \\ Now we
consider an infinitesimally thin non-absorbing dielectric slab,
placed in the plane $z=z_0$ and represented by a non-homogeneous
dielectric constant of equation
\\
\begin{equation}\label{90}
\varepsilon(z) = \varepsilon_0 \left[ \eta \delta (z - z_0) + 1
\right],
\end{equation}
\\
where $\eta$ is a constant depending upon the optical properties
of dielectric media \cite{dutra96}.\\
\begin{picture}(360,280)(-90,30)
\thicklines
  \put(120,70){\framebox(10,200)}
\thinlines
  \put(70,170){\line(1,0){50}}
  \put(180,170){\line(-1,0){50}}
  \put(35,197){\makebox(0,0){$\opa_{L \lambda}(\bk)$}}
  \put(220,197){\makebox(0,0){$\opb_{R \lambda}(\bk)$}}
  \put(35,143){\makebox(0,0){$\opa_{R \lambda}(\bk)$}}
  \put(220,143){\makebox(0,0){$\opb_{L \lambda}(\bk)$}}
  \put(85,180){\makebox(0,0){$\theta$}}
  \put(165,180){\makebox(0,0){$\theta$}}
  \put(60,125){\makebox(0,0){$\bk_+$}}
  \put(190,125){\makebox(0,0){$\bk_-$}}
  \put(40,130){\vector(2,1){76}}
  \put(134,172){\vector(2,1){76}}
  \put(116,172){\vector(-2,1){76}}
  \put(210,130){\vector(-2,1){76}}
  \put(100,170){\oval(20,20)[tl]}
  \put(150,170){\oval(20,20)[tr]}
\end{picture}

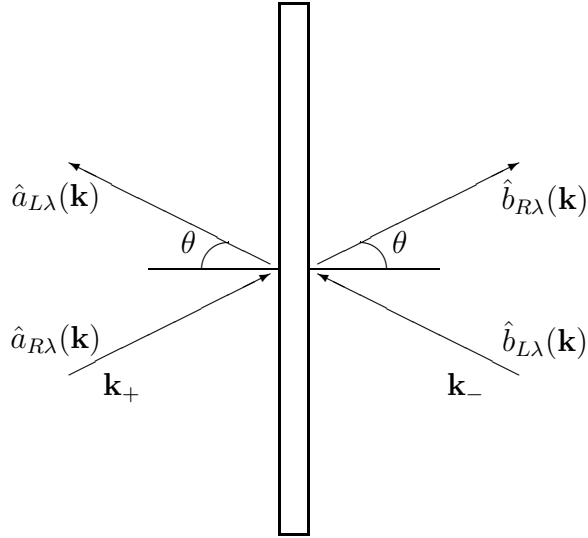
FIG. 2. Schematic representation of the dielectric slab that
transform   {\em Input} operators $\opa_{R \lambda}(\bk)$ and
$\opb_{L \lambda}(\bk)$ in {\em Output} operators $\opa_{L
\lambda}(\bk)$ and $\opb_{R \lambda}(\bk)$.
\\
\\
 The complex reflection and
transmission coefficients $r_\lambda(\bk)$ and $t_\lambda(\bk)$ of
the slab depend upon the incident radiation wave-vector and
polarization and they have the following unitary lossless
properties, valid for all values of $\bk, \, \lambda$:
\\
\begin{equation}\label{100}
\begin{split}
  |r_{ \lambda} (\bk)|^2 + |t_{ \lambda} (\bk)|^2 & = 1,
  \\\\
  t_{ \lambda} (\bk) r_{ \lambda}^* (\bk) + r_{ \lambda} (\bk)
   t_{ \lambda}^* (\bk) & = 0.
\end{split}
\end{equation}
\\
Consider the arrangement in Fig. 2, which shows the interaction of
two fields of wavevectors $\bk_+$ and $\bk_-$ respectively, on a
dielectric slab placed at $z=0$. It is well known
\cite{prasad87,fearn87} that the annihilation operators $\opa_{L
\lambda}(\bk)$ and $\opc_{R \lambda}(\bk)$  associated to Output
modes are related to the annihilation operators $\opa_{R
\lambda}(\bk)$ and $\opc_{L \lambda}(\bk)$  associated to Input
modes, by the $2 \times 2$ matrix
\\
\begin{equation}\label{110}
\begin{pmatrix}
\opa_{L \lambda} (\bk) \\\\ \opc_{R \lambda} (\bk)
\end{pmatrix}  =
 \begin{pmatrix}
   r_{ \lambda} (\bk) & t_{ \lambda} (\bk) \\\\
   t_{ \lambda} (\bk) & r_{ \lambda} (\bk) \
 \end{pmatrix}
\begin{pmatrix}
\opa_{R \lambda} (\bk) \\\\ \opc_{L \lambda} (\bk)
\end{pmatrix},
\end{equation}
\\
whose unitarity is assured by Eqs. (\ref{100}). Now imagine to put
a pair of infinitesimally thin dielectric slabs at $z = \pm l/2$,
to form a planar Fabry-P\'erot cavity, as shown in Fig. 3.
\\
We impose boundary conditions of the form (\ref{110}) both on slab
$1$ and on slab $2$ obtaining respectively,
\\
\begin{equation}\label{120}
\begin{split}
\opb_{R \lambda} (\bk) e^{ i \varphi_R (-l/2)} & = t_{1
\lambda}(\bk) \opa_{R \lambda} (\bk) e^{ i \varphi_R (-l/2)} +
r_{1 \lambda}(\bk) \opb_{L \lambda} (\bk) e^{ i \varphi_L (-l/2)}
,
\\\\ \opa_{L \lambda} (\bk) e^{ i \varphi_L (-l/2)} & = t_{1 \lambda}(\bk)
 \opb_{L \lambda}
(\bk) e^{ i \varphi_L (-l/2)} + r_{1 \lambda}(\bk) \opa_{R
\lambda} (\bk) e^{ i \varphi_R (-l/2)} ,
\end{split}
\end{equation}
\\
and
\\
\begin{equation}\label{130}
\begin{split}
\opc_{R \lambda} (\bk) e^{ i \varphi_R (l/2)} & = t_{2
\lambda}(\bk) \opb_{R \lambda} (\bk) e^{ i \varphi_R (l/2)} + r_{2
\lambda}(\bk) \opc_{L \lambda} (\bk) e^{ i \varphi_L (l/2)} ,
\\\\ \opb_{L \lambda} (\bk) e^{ i \varphi_L (l/2)} & = t_{2 \lambda}(\bk)
 \opc_{L \lambda}
(\bk) e^{ i \varphi_L (l/2)} + r_{2 \lambda}(\bk) \opb_{R \lambda}
(\bk) e^{ i \varphi_R (l/2)},
 \end{split}
\end{equation}
\\
where $\varphi_R(z)$ and $\varphi_L(z)$ are the phases generated
by field propagation inside the cavity.\\
\begin{picture}(480,320)(-50,40)
\thicklines
  \put(110,80){\framebox(10,220)}
  \put(220,80){\framebox(10,220)}
\thinlines
  \put(115,315){\makebox(0,0){{\bf Mirror 1}}}
  \put(225,315){\makebox(0,0){{\bf Mirror 2}}}
  \put(115,65){\makebox(0,0){$z=-l/2$}}
  \put(225,65){\makebox(0,0){$z=l/2$}}
  \put(70,130){\makebox(0,0){$2 \theta$}}
  \put(275,190){\makebox(0,0){$2 \theta$}}
  \put(20,90){\makebox(0,0){$\bk_+$}}
  \put(320,150){\makebox(0,0){$\bk_-$}}
  \put(42,180){\makebox(0,0){$\opa_{L \lambda}(\bk)$}}
  \put(42,110){\makebox(0,0){$\opa_{R \lambda}(\bk)$}}
  \put(170,180){\makebox(0,0){$\opb_{R \lambda}(\bk)$}}
  \put(170,240){\makebox(0,0){$\opb_{L \lambda}(\bk)$}}
  \put(300,170){\makebox(0,0){$\opc_{L \lambda}(\bk)$}}
  \put(300,240){\makebox(0,0){$\opc_{R \lambda}(\bk)$}}
  \put(30,90){\vector(2,1){76}}
  \put(122,136){\vector(2,1){96}}
  \put(234,192){\vector(2,1){76}}
  \put(310,150){\vector(-2,1){76}}
  \put(218,196){\vector(-2,1){96}}
  \put(106,132){\vector(-2,1){76}}
  \put(90,130){\oval(20,20)[l]}
  \put(250,190){\oval(20,20)[r]}
\end{picture}

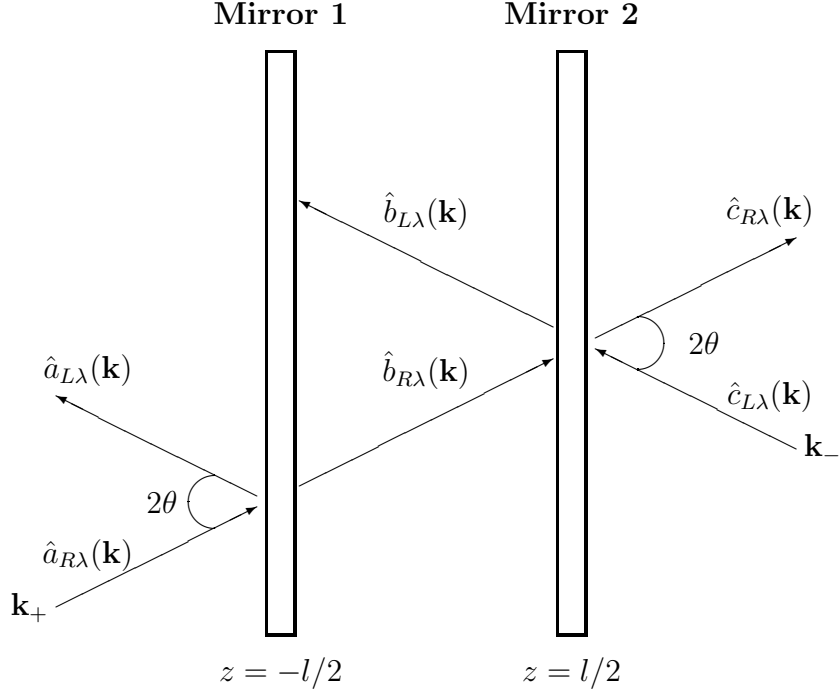
FIG. 3. Schematic representation of a planar Fabry-P\'erot cavity
with notation for {\em Input} operators
 $\opa_{R \lambda}(\bk), \, \opc_{L \lambda}(\bk)$, {\em
Inside} operators
 $\opb_{R \lambda}(\bk), \, \opb_{L \lambda}(\bk)$
  and {\em Outside} operators
  $\opa_{L \lambda}(\bk), \, \opc_{R \lambda}(\bk)$.
\\
\\
 We freely choose the zero
phases in the middle of the cavity, so that $\varphi_F(-z) = -
\varphi_F(z)$, where $F=R, \, L$ and $\varphi_R(z) = -
\varphi_L(z)$. Then we can put
\\
\begin{equation}\label{140}
\varphi_L (\pm l/2) = \mp \delta/2, \qquad \varphi_R (\pm l/2) =
\pm \delta/2,
\end{equation}
\\
where
\\
\begin{equation}\label{150}
 \delta \equiv \frac{\omega}{c} l \cos \theta,
\end{equation}
\\
is half of the phase gained in a double traversal of the cavity
\cite{bornwolf}. From the first of Eqs. (\ref{120}) and the second
of (\ref{130}), we can express the intracavity operators $\opb_{R
\lambda}(\bk)$ and $\opb_{L \lambda}(\bk)$ as
\\
\begin{equation}\label{160}
\opb_{R\lambda} (\bk)  = \opa_{R \lambda}(\bk) I_\lambda(\bk)  +
\opc_{L \lambda}(\bk) J_\lambda'(\bk),
\end{equation}
\begin{equation}\label{170}
\opb_{L\lambda} (\bk) = \opc_{L \lambda}(\bk) J_\lambda(\bk)  +
\opa_{R \lambda}(\bk) I_\lambda'(\bk).
\end{equation}
\\
where we have defined, following Ref. \cite{ley87},
\\
\begin{equation}\label{220}
  D_\lambda(\bk) \equiv 1 - r_{1 \lambda}(\bk) r_{2 \lambda}(\bk)
  \exp (2 i \delta),
\end{equation}
\begin{equation}\label{230}
\begin{split}
 I_{ \lambda} (\bk)  & = t_{1 \lambda}(\bk) /D_\lambda (\bk), \\\
  I_{ \lambda}' (\bk) & = t_{2 \lambda}(\bk) /D_\lambda (\bk),
 \end{split}
\end{equation}
\begin{equation}\label{240}
\begin{split}
 J_{ \lambda} (\bk) & = r_{2 \lambda}(\bk) \exp ({ i \delta})
  I_{ \lambda} (\bk) , \\\
J_{ \lambda}' (\bk) & = r_{1 \lambda}(\bk) \exp ({ i \delta})
 I_{\lambda}' (\bk) .
\end{split}
\end{equation}
\\
Substituting Eqs. (\ref{160})-(\ref{170}) in the second of the
(\ref{120}) and in the first of (\ref{130}), we obtain
\\
\begin{equation}\label{180}
\opc_{R\lambda} (\bk)  = \opa_{R \lambda}(\bk) T_\lambda(\bk)  +
\opa_{L \lambda}(\bk) R_\lambda'(\bk),
\end{equation}
\begin{equation}\label{190}
\opa_{L\lambda} (\bk) = \opc_{L \lambda}(\bk) R_\lambda(\bk)  +
\opa_{R \lambda}(\bk) T_\lambda'(\bk),
\end{equation}
\\
where
\\
\begin{equation}\label{250}
T_{\lambda} (\bk)  = T_{\lambda}' (\bk)  \equiv   \frac{t_{1
\lambda}(\bk) t_{2 \lambda}(\bk)}{D_\lambda(\bk)} ,
\end{equation}
\begin{equation}\label{260}
R_{\lambda} (\bk)   \equiv \frac{r_{1 \lambda}(\bk)  \exp ({- i
\delta} ) + r_{2 \lambda}(\bk) \left[ t_{1 \lambda}^2(\bk) - r_{1
\lambda}^2(\bk) \right] \exp ({ i \delta} ) }{D_\lambda(\bk)},
\end{equation}
\begin{equation}\label{270}
R_{\lambda}' (\bk)   \equiv \frac{r_{2 \lambda}(\bk) \exp ({- i
\delta} ) + r_{1 \lambda}(\bk) \left[ t_{2 \lambda}^2(\bk) - r_{2
\lambda}^2(\bk) \right] \exp ({ i \delta} ) }{D_\lambda(\bk)}.
\end{equation}
\\
The coefficients $R_{\lambda} (\bk)$ and $T_{\lambda}(\bk)$ and
the corresponding primed represent the  reflection and
transmission coefficients of a Fabry-P\'erot cavity as a whole. It
can readily be shown that they satisfy the conditions \cite{ley87}
\\
\begin{equation}\label{280}
\left| R_\lambda (\bk) \right| = \left| R_\lambda' (\bk) \right|,
\end{equation}
\begin{equation}\label{290}
 \left| R_\lambda (\bk) \right|^2 + \left| T_\lambda (\bk)
 \right|^2 =  \left| R_\lambda' (\bk) \right|^2 + \left| T_\lambda' (\bk)
 \right|^2 = 1,
\end{equation}
\begin{equation}\label{300}
R_\lambda^*(\bk) T_\lambda'(\bk) + R_\lambda'(\bk)
T_\lambda^*(\bk) = 0.
\end{equation}
\\
Using   Eqs. (\ref{60}) and (\ref{280})-(\ref{300}) it is not
difficult to show that the Output operators $\opc_{R \lambda}
(\bk)$ and $\opa_{L \lambda} (\bk)$ satisfy canonical commutation
rules
\\
\begin{equation}\label{310}
\begin{split}
 & \left[ \opa_{ L \lambda} (\bk), \,\opa_{L \lambda'}^{ \dag} (\bk')
   \right] =  \left[ \opc_{ R \lambda} (\bk),
    \,\opc_{R \lambda'}^{ \dag} (\bk') \right]
   =   \delta_{\lambda {\lambda'}} \delta (\bk - \bk') \\\\
&  \left[ \opa_{ L \lambda} (\bk), \,\opc_{R \lambda'}^{ \dag}
(\bk')
   \right] =  \left[ \opc_{ R \lambda} (\bk),
    \,\opa_{L \lambda'}^{ \dag} (\bk') \right] =  0,
\end{split}
\end{equation}
\\
while the intracavity operators $\opb_{R \lambda} (\bk)$ and
$\opb_{L \lambda} (\bk)$ satisfy anomalous commutation rules
\cite{ueda94,barnett96}
\\
\begin{equation}\label{320}
\begin{split}
  \left[ \opb_{ R  \lambda} (\bk), \,\opb_{R  \lambda'}^{ \dag} (\bk')
   \right]
  & = \delta_{\lambda {\lambda'}} \delta (\bk - \bk')
  \frac{1 - \left| r_{1 \lambda}(\bk) r_{2 \lambda}(\bk)
 \right|^2 }
  { \left| 1 - r_{1 \lambda}(\bk) r_{2 \lambda}(\bk)
  e^{2 i \delta} \right|^2} \\\\
  & =   \left[ \opb_{ L  \lambda} (\bk), \,\opb_{L  \lambda'}^{ \dag} (\bk')
   \right],
\end{split}
\end{equation}
\begin{equation}\label{340}
\begin{split}
  \left[ \opb_{ L  \lambda} (\bk), \,\opb_{R \lambda'}^{ \dag} (\bk')
   \right]
  &  = \delta_{\lambda {\lambda'}} \delta (\bk - \bk')
  \frac{ r_{2 \lambda}(\bk)e^{  i \delta}
  [ 1 - \left| r_{1 \lambda}(\bk) \right|^2 ]
  +
r_{1 \lambda}^* (\bk)e^{ - i \delta}
  [ 1 - \left| r_{2 \lambda}(\bk) \right|^2 ]
   }
  { \left| 1 - r_{1 \lambda}(\bk) r_{2 \lambda}(\bk)
  e^{2 i \delta} \right|^2} \\\\
 &  =   \left[ \opb_{ R  \lambda} (\bk), \,\opb_{L  \lambda'}^{\dag} (\bk')
   \right]^*.
\end{split}
\end{equation}
\\
These equations are the three-dimensional generalization of Eqs.
(9) given in Ref. \cite{barnett96} for one-dimensional fields.
\\
Because of the presence of the cavity, the vector potential is now
written as
\\
\begin{equation}\label{350}
 \opAv^+ (\br,t)   =   \int d \bk  \left( \frac{\hbar
}{16 \pi^3 \varepsilon_0  \omega } \right)^{1/2} \sum_{\lambda =
1,2} \mathbf{{F}}_\lambda (\bk, \br)  \exp (- i \omega t) ,
\end{equation}
\\
where
\\
\begin{equation}\label{360}
\mathbf{{F}}_\lambda (\bk, \br) =
\begin{cases}
\opa_{R \lambda}(\bk) \bepsilon_\lambda(\bk_+) e^{i \bk_+ \cdot
\br}  + \opa_{L \lambda}(\bk) \bepsilon_\lambda(\bk_-) e^{i \bk_-
\cdot \br} \, , & \quad -\infty<z<l/2 \, ,
\\\\
\opb_{R \lambda}(\bk) \bepsilon_\lambda(\bk_+) e^{i \bk_+ \cdot
\br}  + \opb_{L \lambda}(\bk) \bepsilon_\lambda(\bk_-) e^{i \bk_-
\cdot \br} \, ,
   & \quad -l/2<z<l/2 \, ,  \\\\
\opc_{R \lambda}(\bk) \bepsilon_\lambda(\bk_+) e^{i \bk_+ \cdot
\br}  + \opc_{L \lambda}(\bk) \bepsilon_\lambda(\bk_-) e^{i \bk_-
\cdot \br} \, , & \quad l/2<z<\infty \, .
  \end{cases}
\end{equation}
\\
We note that using different pairs of annihilation operators for
each region of space delimited by  the cavity, as in Eqs.
(\ref{350}-\ref{360}), we obtain a free-field like representation
for all space inside and outside the cavity, but the canonical
commutation rules are lost for intracavity operators, as shown by
Eqs. (\ref{320}-\ref{340}). Conversely if we choose the mode
functions for example as in Ref. \cite{FDMI91}, we lost free-field
like representation, but we obtain canonical commutation rules for
annihilation and creation operators in whole space.
 It is easy to show that our result agrees with that of Ref. \cite{FDMI91},
  indeed  defining
\\
\begin{equation}\label{370}
\begin{split}
 \opa_{R \lambda} (\bk) & \equiv \opa_{\bk \lambda} , \\\
\opc_{L \lambda} (\bk) & \equiv \opa_{\bk \lambda}' ,
\end{split}
\end{equation}
\\
and substituting Eqs. (\ref{160}-\ref{170}) and
(\ref{180}-\ref{190}) into  Eqs. (\ref{360}), by a straightforward
calculation, we obtain
\\
\begin{equation}\label{380}
\mathbf{{F}}_\lambda (\bk, \br) = \bepsilon (\bk, \lambda) \left(
U_{\bk \lambda} \opa_{\bk \lambda} + U_{\bk \lambda}' \opa_{\bk
\lambda}'  \right), \qquad - \infty < z < + \infty,
\end{equation}
\\
where now the mode function $U_{\bk \lambda}$ are defined
differently on the three regions of the space, as shown in the
tables (2.6) and (2.7) of Ref. \cite{FDMI91}.
\section{One-dimensional formulation}
In this work we are interested to derive Input-Output relations
for a single transverse mode of the cavity, having finite
cross-section area $\mathcal{A}$ orthogonal to the $z$ axis which,
in fact, depend upon the geometrical and transmitting properties
of the cavity itself \cite{aiello95,ujihara91}. Then, following
Ref. \cite{blow90}, we impose periodic boundary conditions on both
directions $x$ and $y$, so that the corresponding components of
the wave-vector are restricted to the discrete values
\\
\begin{equation}\label{d.4}
  \dis{k_x =  \frac{2 \pi}{\mathcal{A}^{1/2}}n_x, \quad k_y =
  \frac{2 \pi}{\mathcal{A}^{1/2}}n_y, \quad
  \text{where} \quad n_x, n_y = 0, \pm 1, \, \pm 2, \, \ldots}
\end{equation}
\\
while $k_z$ persists to be continuous positive variable. Then the
following conversions are required:
\\
\begin{equation}\label{d.5}
\dis{ \int d \bk \rightarrow  \frac{(2 \pi)^2}{\mathcal{A}}
\sum_{k_x,k_y} \int_0^\infty dk_z },
\end{equation}
\begin{equation}\label{d.9}
 \dis{ \delta^3(\bk - \bk') \rightarrow
  \frac{\mathcal{A}}{(2 \pi)^2} \delta_{\bn \bn'} \delta (k_z - k_z')
  },
\end{equation}
\begin{equation}\label{d.7}
\begin{split}
 \opa_{ R \lambda} (\bk) & \rightarrow
  (2 \pi /\mathcal{A}^{1/2})^{-1} \opa_{ R \lambda}
   (2 \pi /\mathcal{A}^{1/2} \, \bn, k_z) ,
    \\\
   \opc_{ L \lambda} (\bk) & \rightarrow
  (2 \pi /\mathcal{A}^{1/2})^{-1} \opc_{ L \lambda}
   (2 \pi /\mathcal{A}^{1/2} \, \bn, k_z) ,
\end{split}
\end{equation}
\\
where $\bn \equiv (n_x,n_y)$. Now we fix a linear polarization
parallel to the $x$ axis,  consider only field excitations with
$n_x = n_y =0$ and define
\\
\begin{equation}\label{d.13}
\begin{split}
  \opa_{ R \lambda} (2 \pi /\mathcal{A}^{1/2} \, \bn, k_z)
   \Big{|}_{\bn = \mathbf{0}}
& \equiv c^{1/2} \opa_R( \omega ) , \\\
  \opc_{ L \lambda} (2 \pi /\mathcal{A}^{1/2} \, \bn, k_z)
  \Big{|}_{\bn = \mathbf{0}} & \equiv c^{1/2} \opc_L( \omega ) ,
\end{split}
\end{equation}
\\
where we have defined  $k_z \equiv k \equiv \omega /c$.  These
operators have the commutators
\\
\begin{equation}\label{d.14}
 \begin{split}
\left[ \opa_R( \omega ), \, \opa_R^\dag( \omega' ) \right] =
\left[\opc_L( \omega ), \, \opc_L^{\dag}( \omega' ) \right] & =
\delta(\omega-\omega'), \\\\ \left[ \opa_R( \omega ), \,
\opc_L^{\dag}( \omega' ) \right] = \left[ \opc_L( \omega ), \,
\opa_R^\dag( \omega ') \right] & = 0.
 \end{split}
\end{equation}
\\
Until now,  we have limited our attention only to incident
operators, however repeating the same procedure for operators
$\opb_{R \lambda}(\bk)$, $\opb_{R \lambda}(\bk)$ and $\opa_{L
\lambda}(\bk)$, $\opc_{R \lambda}(\bk)$, we define
straightforwardly the corresponding one-dimensional operators
$\opb_{R }(\omega)$, $\opb_{L}(\omega)$ and $\opa_{L }(\omega)$,
$\opc_{R }(\omega)$.\\ Now we assemble the six operators $\opa_{R
}(\omega)$, $\opc_{L}(\omega)$, $\opb_{R }(\omega)$,
$\opb_{L}(\omega)$ and $\opa_{L }(\omega)$, $\opc_{R }(\omega)$ in
three doublets defining
\\
\begin{equation}\label{ciao6.1}
\hat{\mathbf{a}}(\omega) = \left( \begin{array}{c}
  \opa_1(\omega) \\\\
  \opa_2(\omega)
\end{array}
\right)  \equiv \left( \begin{array}{c}
  \opa_R(\omega) \\\\
  \opc_L(\omega)
\end{array}
\right),
\end{equation}
\begin{equation}\label{ciao6.2}
\hat{\mathbf{b}}(\omega) = \left( \begin{array}{c}
  \opb_1(\omega) \\\\
  \opb_2(\omega)
\end{array}
\right)  \equiv \left( \begin{array}{c}
  \opb_R(\omega) \\\\
  \opb_L(\omega)
\end{array}
\right),
\end{equation}
\begin{equation}\label{ciao6.3}
\hat{\mathbf{c}}(\omega) = \left( \begin{array}{c}
  \opc_1(\omega) \\\\
  \opc_2(\omega)
\end{array}
\right)  \equiv \left( \begin{array}{c}
  \opc_R(\omega) \\\\
  \opa_L(\omega)
\end{array}
\right).
\end{equation}
\\
Their geometric meaning is illustrated in Fig. 4.\\
  \begin{picture}(400,200)(-20,80)
\thinlines
  \put(40,190){\vector(1,0){80}}
    \put(80,200){\makebox(0,0){$
    \opa_1 (\omega)$}}
  \put(160,190){\vector(1,0){80}}
    \put(200,200){\makebox(0,0){$
    \opb_1 (\omega)$}}
  \put(280,190){\vector(1,0){80}}
    \put(320,200){\makebox(0,0){$
    \opc_1 (\omega)$}}
  \put(120,140){\vector(-1,0){80}}
   \put(80,150){\makebox(0,0){$
   \opc_2 (\omega)$}}
  \put(240,140){\vector(-1,0){80}}
   \put(200,150){\makebox(0,0){$
    \opb_2 (\omega)$}}
  \put(360,140){\vector(-1,0){80}}
   \put(320,150){\makebox(0,0){$
    \opa_2 (\omega)$}}
  \put(140,250){\makebox(0,0){{\bf Mirror 1}}}
  \put(260,250){\makebox(0,0){{\bf Mirror 2}}}
  \put(140,90){\makebox(0,0){$z = -l/2$}}
  \put(260,90){\makebox(0,0){$z = l/2$}}
 \thicklines
 \put(135,110){\framebox(10,110)}
 \put(255,110){\framebox(10,110)}
\end{picture}

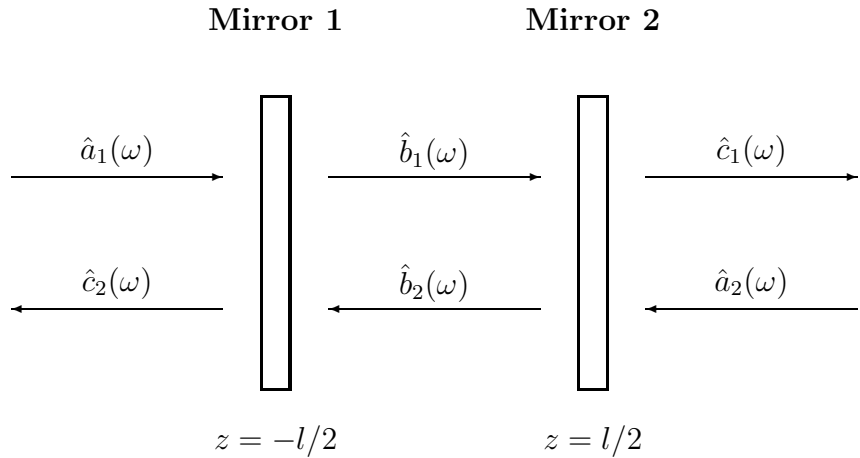
FIG. 4. Schematic representation of a one-dimensional
Fabry-P\'erot cavity with notation for {\em Input} operators
$\opa_i(\omega), \, i=1,2$, {\em Inside} operators
$\opb_i(\omega), \, i=1,2$ and {\em Outside} operators
$\opc_i(\omega), \, i=1,2$.
\\
\\
Using Eqs. (\ref{160}-\ref{170}) in one-dimensional form, it is
not difficult to show with a straightforward calculation, that the
{\em Inside} operators $\opb_i(\omega), \; i=1,2$ are related to
{\em Incident} operators $\opa_i(\omega), \; i=1,2$ by the
relations
\\
\begin{equation}\label{ciao6}
  \hat{\mathbf{b}}(\omega)  = \mathcal{B}(\omega)
    \hat{\mathbf{a}}(\omega),
\end{equation}
\\
where the matrix elements $B_{i j}(\omega)$ of
$\mathcal{B}(\omega)$ are given by:
\\
\begin{equation}\label{bo1}
\left\{
\begin{split}
B_{i i}(\omega) & = t_i(\omega) / D(\omega) , \\\\ B_{i
j}(\omega)\Big{|}_{i \neq j} & = r_i(\omega) \exp (i \omega l /c)
B_{j j }(\omega) ,
\end{split}
\right.  \qquad i= 1,2, \; j= 1,2 \; ,
\end{equation}
\\
where
\\
\begin{equation}\label{bo2}
 D(\omega) \equiv 1 - r_1(\omega) r_2 (\omega) \exp (2 i \omega l
 /c).
\end{equation}
\\
Similarly, the {\em Outside} operators $\opc_i(\omega), \, i=1,2$
are related to the {\em Incident} operators by
\\
\begin{equation}\label{bo3}
  \hat{\mathbf{c}}(\omega)  = \mathcal{C}(\omega)
    \hat{\mathbf{a}}(\omega),
\end{equation}
\\
where the matrix elements $C_{i j}(\omega)$ $(  i, j= 1,2)$ of
$\mathcal{C}(\omega)$ are given by:
\\
\begin{equation}\label{bo4}
\left\{
\begin{split}
C_{i i}(\omega) & = t_1(\omega) t_2(\omega) / D(\omega) , \\\\
C_{i j}(\omega) \Big{|}_{i \neq j} & =\frac{ r_j(\omega) \exp (-i
\omega l /c) + r_i(\omega) \exp [i \omega l /c + 2 i \arg
t_j(\omega)] }{D(\omega)} .
\end{split}
\right.
\end{equation}
\\
Using Eqs. (\ref{100})-(\ref{bo4}) it is easy to show that the
matrix $\mathcal{C} (\omega)$ is unitary:
\\
\begin{equation}\label{bo5}
\mathcal{C} (\omega) \cdot \mathcal{C}^\dag (\omega) =
\mathcal{C}^\dag (\omega) \cdot \mathcal{C} (\omega) = \mathbb{I}
.
\end{equation}
\\
This leads, with Eqs. (\ref{d.14})-(\ref{bo3}), to the result
\\
\begin{equation}\label{e.1}
\left[ \opa_i(\omega),\opa_j^\dag(\omega') \right]= \left[
\opc_i(\omega),\opc_j^\dag(\omega') \right] = \delta_{i j}
\delta(\omega- \omega'), \qquad i,j = 1,2.
\end{equation}
\\
For {\em Inside} operators the commutation rules are found to be
\\
\begin{equation}\label{ciao4}
\begin{split}
 \left[ \opb_i(\omega),\opb_j^\dag (\omega') \right]
  & = \left[ \mathcal{B}(\omega) \mathcal{B}^\dag(\omega') \right]_{i j}
    \delta(\omega-\omega')
  \\\\
& \equiv {G}_{i j}(\omega) \delta(\omega-\omega') ,
\end{split}
\end{equation}
\\
where, for construction, $\mathcal{G}(\omega) = \mathcal{G}^\dag
(\omega)$, being ${G}_{i j}(\omega) \equiv
[\mathcal{G}(\omega)]_{i j}$, and
\\
\begin{equation}\label{bo8}
\left\{
\begin{split}
G_{1 1}(\omega) & = \frac{1 - |r_1(\omega)
r_2(\omega)|^2}{D(\omega)}  = G_{2 2}(\omega) ,
\\\\ G_{1 2}(\omega)  & = \frac{
r_2(\omega)\exp(i \omega l /c) [1 -  |r_1(\omega) |^2 ] +
r_1^*(\omega)\exp(-i \omega l /c) [1 -  |r_2(\omega) |^2 ]
}{D(\omega)} .
\end{split}
\right.
\end{equation}
\\
Noting that $\mathrm{Det}[\mathcal{B} (\omega)]  \neq 0$ for
$t_1(\omega) \neq 0$ and $t_2(\omega) \neq 0$, we can invert it to
express the relation between Inside and Outside operators as
\\
\begin{equation}\label{ciao6}
\begin{split}
  \hat{\mathbf{c}}(\omega) & = \mathcal{C}(\omega) \mathcal{B}^{-1}(\omega)
    \hat{\mathbf{b}}(\omega) \\\\
    & \equiv \mathcal{M}(\omega)
    \hat{\mathbf{b}}(\omega),
\end{split}
\end{equation}
\\
where
\\
\begin{equation}\label{d.91}
\begin{split}
\mathcal{M}(\omega)
 & =  \begin{pmatrix}
  \dis{\frac{1}{t_2^*(\omega)}} & \dis{\frac{r_2(\omega)}{t_2(\omega)}
   \exp{(- i \omega l/c)}
  }\\\\
 \dis{\frac{r_1(\omega)}{t_1(\omega)} \exp{(- i \omega l/c)}} & \dis{
 \frac{1}{t_1^*(\omega)}}
 \end{pmatrix}.
\end{split}
\end{equation}
\\
It should be noted that the generic $2 \times 2$ matrix that
represents a  Beam-Splitter must be {\bf unitary} to preserve the
canonical bosonic relations of commutation  for both {\em Input}
and {\em Output} operators; in our case $\mathcal{M}(\omega)$ is
not unitary at all. In fact, while {\em Output} operators
$\opc_i(\omega)$ satisfy the canonical relations (\ref{e.1}), {\em
Input}  operators $\opb_i(\omega)$ satisfy  the relations
(\ref{ciao4}), that is the so-called anomalous relations of
commutation. Using Eqs. (\ref{ciao6}-\ref{d.91}) we obtain
\\
\begin{equation}\label{ciao7}
\mathcal{M}^\dag(\omega) \mathcal{M}(\omega) =
\mathcal{G}^{-1}(\omega),
\end{equation}
\\
or, equivalently,
\\
\begin{equation}\label{ciao8}
\mathcal{M}(\omega) \mathcal{G}(\omega) \mathcal{M}^\dag(\omega) =
\mathbb{I}.
\end{equation}
\\

Because of the non-unitarity of $\mathcal{M}(\omega)$, the
photon-number operator is not conserved on a single mode. In fact
from Eq.(\ref{ciao6}-\ref{ciao7}) we obtain
\\
\begin{equation}\label{ciao8.1}
  \hat{\mathbf{c}}^\dag(\omega) \cdot   \hat{\mathbf{c}}(\omega)
  =   \hat{\mathbf{b}}^\dag(\omega) \cdot \mathcal{G}^{-1}(\omega)
 \cdot  \hat{\mathbf{b}}(\omega) \neq \hat{\mathbf{b}}^\dag(\omega)
 \cdot  \hat{\mathbf{b}}(\omega) .
\end{equation}
\\
 Anyway, because we are working with linear transformations,
the most general bilinear form in {\em Inside} creation operators
$\opb_i^\dag(\omega)$ is still bilinear in {\em Outside} creation
operators $\opc_i^\dag(\omega)$:
\\
\begin{equation}\label{d.92.3}
\begin{split}
 \gamma_{11} \opc_1^\dag \opc_1^\dag & + \gamma_{12} \opc_1^\dag \opc_2^\dag
+ \gamma_{21} \opc_2^\dag \opc_1^\dag
  +
\gamma_{22} \opc_2^\dag \opc_2^\dag
\\\\ = &
 \beta_{11} \opb_1^\dag \opb_1^\dag  + \beta_{12} \opb_1^\dag \opb_2^\dag
+ \beta_{21} \opb_2^\dag \opb_1^\dag
  +
\beta_{22} \opb_2^\dag \opb_2^\dag.
\end{split}
\end{equation}
\\
Because the first of the two forms applied to free space generates
a two-photon state of the electromagnetic field, the equality
ensures the same for the second one. Therefore, if it is possible
to associate to the most general two-photon state generated by
{\em Inside} operators a state with two photons physically
 generated {\bf inside} the cavity, then,
from the relations between {\em Input} and {\em Output} operators,
we can obtain information about the field outside the cavity, that
is the actual object of measurement. In the following section we
study how we can do this.

\section{States of the Field}
 We define the states generated by the linear and bilinear forms of
{\em Inside} operators as:
\\
\begin{equation}\label{e.2}
\begin{array}{rcl}
\opb_i^\dag(\omega) |0\ra & \equiv & |  F_i (\omega) ; \ing  \ra,
\\\\
\opb_i^\dag(\omega) \opb_j^\dag(\omega') |0\ra & \equiv & |  F_i
(\omega),   F_j (\omega')  ; \ing  \ra,
\end{array}
\end{equation}
\\
where $F_i (\omega)$ is a label which depends on continuous
variable $\omega$ and on discrete variable $i=1,2$. Because of
(\ref{ciao4}) these states are not orthogonal:
\\
\begin{equation}\label{e.3}
 \la  F_i (\omega) ; \ing  |   F_j (\omega') ; \ing  \ra =
 G_{i
j} (\omega)  \delta(\omega- \omega'),
\end{equation}
\\
and
\\
\begin{equation}\label{e.4}
\begin{split}
 \la  F_i (\omega_1),F_j (\omega_2) ; \ing  | &   F_k (\omega_3),
  F_l (\omega_4) ; \ing  \ra = \\\\   &
 G_{ik} (\omega_1)  G_{j l} (\omega_2) \delta(\omega_1 - \omega_3)
 \delta(\omega_2 - \omega_4) \\\\  & +   G_{i l} (\omega_1)
   G_{j k} (\omega_2)
 \delta(\omega_2 - \omega_3)
 \delta(\omega_1 - \omega_4).
 \end{split}
\end{equation}
\\
From Eqs. (\ref{e.3}-\ref{e.4}) we note that anomalous commutation
rules, represented by the $2 \times 2$ hermitian matrix
$\mathcal{G}(\omega)$, form a metric in the two-dimensional
Hilbert space generated by {\em Inside} operators
$\opb_i(\omega)$. For example if we write the most general
one-photon state created by {\em Inside} operators as
\\
\begin{equation}\label{unf.1}
| \phi \ra = \sum_{i =1}^2 \int d \omega \, K_i (\omega) |
F_i(\omega); \ing \ra,
\end{equation}
\\
where $K_i (\omega)\in \mathbb{C}$, then its norm is
 \\
\begin{equation}\label{unf.6}
\begin{split}
  \la \phi | \phi \ra &  = \sum_{i,j}^{1,2} \int d \omega \,
  K_i^*(\omega) G_{i j }(\omega) K_j(\omega) \\\\
  &    = \int d \omega \, {\mathbf{K}}^\dag (\omega)
   \cdot \mathcal{G} (\omega) \cdot \mathbf{K} (\omega),
\end{split}
\end{equation}
\\
where it is clear the metric-like role of matrix
$\mathcal{G}(\omega)$. \\ As before, we can define the states
generated by the linear and bilinear forms of {\em Outside}
operators as
\\
\begin{equation}\label{e.5}
\begin{array}{rcl}
\opc_i^\dag(\omega) |0\ra & \equiv & |  F_i (\omega) ; \out  \ra ,
\\\\
\opc_i^\dag(\omega) \opc_j^\dag(\omega') |0\ra & \equiv & |  F_i
(\omega),   F_j (\omega')  ; \out  \ra ,
\end{array}
\end{equation}
\\
which, using Eq. (\ref{ciao6}), can be written in terms of {\em
Inside} operators:
\\
\begin{equation}\label{e.6}
\begin{array}{rcl}
 | F_i (\omega) ; \out \ra & \equiv & \dis{\sum_k^{1,2}} M_{i k}^*(\omega)
  |  F_i (\omega) ; \ing  \ra,
\\\\
 | F_i (\omega), F_j (\omega') ; \out \ra & \equiv &
 \dis{\sum_{k,l}^{1,2}}
 M_{i k}^* (\omega) M_{j l}^*  (\omega')
  | F_k (\omega), F_l (\omega') ; \ing \ra,
\end{array}
\end{equation}
\\
where $M_{i j}  (\omega) \equiv [\mathcal{M}]_{i j}$.  Of course
these state are orthonormal.
\\
It is possible to make number-states also for a continuous
distribution of modes, following  Blow{\em et al.}'s method
\cite{blow90}. On this purpose we define two operators
$\opC_i(\eta)$ as
\\
\begin{equation}\label{e.8}
\opC_i(\eta) = \int d \omega \, \eta_i^*(\omega) \opc_i(\omega),
\qquad i = 1,2 \, ,
\end{equation}
\\
where $\eta_i(\omega)$ are two arbitrary complex functions which
satisfy the normalization condition.
\\
\begin{equation}\label{c6}
 \int d \omega | \eta_i(\omega)|^2 = 1, \qquad i = 1,2 .
\end{equation}
\\
It is easy to verify that these new operators obey the following
commutation relations:
\\
\begin{equation}\label{c7}
\left[ \opC_i(\eta),\opC_j^\dag(\eta) \right] = \delta_{i j},
\end{equation}
\\
and therefore they can be used to make the number-states by the
usual method,
\\
\begin{equation}\label{c7.1}
| F_i^n(\eta) ; \out \ra = (n !)^{-1/2} [ \opC_i^\dag(\eta) ]^n |
0 \ra.
\end{equation}
\\
Using (\ref{c6}) you can verify that the states $|F_i^n(\eta) ;
\out \ra $ are correctly normalized:
\\
\begin{equation}\label{c7.2}
\la  F_i^n(\eta) ; \out| F_j^m(\eta) ; \out \ra = \delta_{i j}
\delta_{n m}.
\end{equation}
\\
The construction of number-states is more problematic for {\em
Inside} operators. As before we define
\\
\begin{equation}\label{c8}
\opB_i(\xi) = \int d \omega \, \xi_i^*(\omega) \opb_i(\omega),
\qquad i = 1,2 \; ,
\end{equation}
\\
where $\xi_i (\omega)$ are two complex arbitrary functions which
can be chosen to satisfy the four conditions
\\
\begin{equation}\label{c9}
\left[ \opB_i(\xi),\opB_j^\dag(\xi) \right] = \int d \omega \,
\xi_i^*(\omega) G_{i j} (\omega) \xi_j(\omega) \equiv {\Gamma}_{i
j}(\xi),
\end{equation}
\\
where ${\Gamma}_{i j}(\xi)$ is a given matrix. Suppose to set
${\Gamma}_{i j}(\xi) = \delta_{i j}$. Because
 $\mathcal{\Gamma}(\xi)$ is hermitian by construction,
Eq (\ref{c9}) corresponds to $2 \oplus 2$ conditions, two real and
a complex one , which the two complex arbitrary functions $\xi_i
(\omega) \; (i = 1,2)$
 must satisfy.
The $\xi_i$ modules can be determined imposing ${\Gamma}_{i
i}(\xi) = 1$. In fact, given an arbitrary function
$\bar{\xi}(\omega)$ such as
\\
\begin{equation}\label{c10}
\int d \omega  |\bar{\xi}(\omega)|^2 = 1,
\end{equation}
\\
if $G_{ii}(\omega) \neq 0$, that is if $|r_1(\omega)
r_2(\omega)|^2 \neq 1$, we can write $\xi_i(\omega)$ as
\\
\begin{equation}\label{c11}
\xi_i(\omega)= \frac{|\bar{\xi}(\omega)| e^{i
\phi_i(\omega)}}{\sqrt{G_{ii}(\omega)}}, \qquad i = 1,2 \; ,
\end{equation}
\\
where $\phi_i(\omega)$ is an arbitrary phase, and we can obtain
$\Gamma_{ii}(\xi) = 1$. Phases are still arbitrary, but in
off-diagonal elements of $\mathcal{\Gamma}(\xi)$ there is only the
phase difference between $\xi_1(\omega)$ and $\xi_2(\omega)$ which
is not sufficient, by itself, to satisfy the two requested
conditions. In fact, you can see that only proper linear
combinations of {\em Inside} operators can generate canonical
commutation relations. Consider  the unitary matrix $\mathcal{U}
(\omega)$ that makes $\mathcal{G}(\omega)$ diagonal
\\
\begin{equation}\label{unf.8}
\mathcal{U}^\dag (\omega) \cdot \mathcal{G}(\omega) \cdot
\mathcal{U} (\omega) = \mathcal{D} (\omega),
\end{equation}
\\
where the diagonal matrix $\mathcal{D}(\omega)$ has elements $D_{i
j} (\omega) = \lambda_i (\omega) \delta_{i j}$,  being
$\lambda_i(\omega), \; i = 1,2 $ the two $\mathcal{G} (\omega)$'s
eigenvalues,
\\
\begin{equation}\label{unf.9}
\lambda_i (\omega) = G_{1 1}(\omega) -(-1)^i |G_{1 2}(\omega)|,
\quad i = 1,2 \; .
\end{equation}
\\
Then if we define the operators $\opd_i(\omega)$ as
\\
\begin{equation}\label{unf.10}
  \hat{\bd}(\omega) \equiv \mathcal{U}^\dag (\omega)
   \hat{\mathbf{b}}(\omega),
\end{equation}
\\
you can see that they satisfy the following ``quasi-canonical''
relations:
\\
\begin{equation}\label{unf.11}
\bigl[ \opd_i(\omega), \opd_j^\dag (\omega') \bigr]= \lambda_i
(\omega) \delta_{i j} \delta(\omega - \omega').
\end{equation}
\\
If we want to obtain fully canonical relations it is necessary to
break the unitariety of the relation between operators
$\opd_i(\omega)$ and $\opb_i(\omega)$ introducing the matrix
$\mathcal{E} (\omega)$ with elements $E_{i j} = (\lambda_i)^{-1/2}
\delta_{i j}$ and to  define the operators $\opb_i'(\omega)$ as
\\
\begin{equation}\label{unf.12}
  \hat{\mathbf{b}}'(\omega) \equiv  \mathcal{E}(\omega)
   \hat{\mathbf{d}}(\omega).
\end{equation}
\\
In fact an easy calculation shows that
\\
\begin{equation}\label{unf.13}
\bigl[ \opb_i'(\omega), \opb_j^{' \dag} (\omega') \bigr]  =
 \delta_{i j} \delta(\omega - \omega'),
\end{equation}
\\
where, using Eq. (\ref{unf.12}),
\\
\begin{equation}\label{unf.7}
\opb_j'(\omega )  \equiv \frac{1}{\sqrt{2 \lambda_j(\omega) }}
\left[ e^{i \phi(\omega)/2} \opb_1(\omega) -(-1)^j e^{-i
\phi(\omega)/2} \opb_2(\omega) \right]  \qquad j = 1,2 \; ,
\end{equation}
\\
and $ \phi(\omega) = \arg [G_{1 2}(\omega)]$ is the relative phase
of the two components of $\mathcal{G}(\omega)$ eigenvectors.\\ Of
course by means of these operators we could make orthonormal {\em
Input} number-state but they will result to be associated to
functions $\sin(\omega z / c)$ and
 $\cos(\omega z/c)$ and, as a consequence, the free
field-like representation will be lost.

 However it is still
possible to write the most general two-photon state generated by
{\em Inside} operators $\opb_i(\omega)$ as,
\\
\begin{equation}\label{e.7}
| \psi \ra = \sum_{i,j}^{1,2} 
  \int d
\omega \int d \omega'
 K_{ij} (\omega, \omega')
 | F_i (\omega), F_j (\omega') ; \ing \ra,
\end{equation}
\\
where, by construction, the matrix $\mathcal{ K}(\omega, \omega')$
of elements  $K_{ij} (\omega, \omega')$, satisfy
\\
\begin{equation}\label{e.8.1}
\mathcal{ K}(\omega, \omega') =  \mathcal{K}^\mathrm{T}(\omega',
\omega),
\end{equation}
\\
where ``T'' indicate transposition. In fact the matrix $\mathcal{
K}(\omega, \omega')$ is fixed by the emission process inside the
cavity.\\ In a simpler way, a two-photon state generated by {\em
Outside} operators (\ref{e.8}), can be written as
\\
\begin{equation}\label{e.9}
\begin{split}
& | F_a (\eta), F_b (\eta) ; \out \ra = (2^{-1/2})^{\delta_{a b}}
\opC_a^\dag (\eta) \opC_b^\dag (\eta) | 0 \ra
\\\\ & \; \; \; \; \; \; \;
 = (2^{-1/2})^{\delta_{a b}} \int d
\omega \int d \omega' \eta_a (\omega) \eta_b (\omega')
 | F_a (\omega), F_b (\omega') ; \out \ra,
 \end{split}
\end{equation}
\\
where $ (a,b = 1,2)$. In the next section we will show how to
express this state by {\em Inside} states.
\section{Two-photon states probability distributions}
It is well know that the inverse of photon mean flight time in a
planar Fabry-P\'erot cavity is given by \cite{knoll}
\\
\begin{equation}\label{e.8.2}
  \gamma_{\mathrm{cav}} \cong\frac{c}{l}\frac{1 - |r_1(\omega) r_2(\omega)|}
  {2 |r_1(\omega) r_2(\omega)|^{1/2}}.
\end{equation}
\\
Now we consider the spontaneous emission of a pair of two
identical atoms or molecules within the microcavity
\cite{deangelis99}. Let $\gamma_{\mathrm{atom}}$ the single atomic
decay rate. In the atom-dominate decay regime (that is when
$\gamma_{\mathrm{cav}} \ll \gamma_{\mathrm{atom}}$ \cite{exter96}
) , for  $1/ \gamma_{\mathrm{atom}} \ll t \ll 1/
\gamma_{\mathrm{cav}}$ the electromagnetic field can be found in a
state like to $| \psi\ra$. If the matrix  $\mathcal{M}(\omega)$
should be unitary, we should calculate easily, as in the quantum
theory of a lossless beam-splitter \cite{campos89}, the
probability distribution of photon number-states outside the
cavity. This is not our case, however we will see that after
renormalization of the state $| \psi\ra$ coefficients  $K_{i
j}(\omega, \omega')$,
 it is possible to obtain significant
results. For this, we calculate the probability amplitude to find
the electromagnetic field, represented by the state $|\psi\ra$
within the cavity, in the outside state $| F_a (\eta), F_b (\eta)
; \out \ra $, $ (a,b = 1,2)$. It is simple to show with the use of
Eqs. (\ref{e.6}) and (\ref{unf.7})-(\ref{e.9}), that result is
\\
\begin{equation}\label{e.10}
\begin{array}{rcl}
 \la F_a (\eta), F_b (\eta) ; \out | \psi \ra &  = & \dis{  2
(2^{-1/2})^{\delta_{a b}} \int d \omega \int d \omega' \eta_a
(\omega) \eta_b (\omega') } \\\\ & & \times \; \left[
\mathcal{M}(\omega) \cdot \mathcal{G}(\omega) \cdot
\mathcal{K}(\omega, \omega') \cdot
\mathcal{G}^{\mathrm{T}}(\omega') \cdot
\mathcal{M}^{\mathrm{T}}(\omega') \right]_{a b}.
\end{array}
\end{equation}
\\
By a lengthy but straightforward calculation, it is simple to show
that
\\
\begin{equation}\label{e.12}
\left[ \mathcal{M}(\omega) \cdot \mathcal{G}(\omega) \cdot
\mathcal{K}(\omega, \omega') \cdot
\mathcal{G}^{\mathrm{T}}(\omega') \cdot
\mathcal{M}^{\mathrm{T}}(\omega') \right]_{a b} \equiv P_{a
b}(\omega, \omega'),
\end{equation}
\\
where we have defined the $2 \times 2$ matrix elements $P_{a
b}(\omega, \omega')$ as
\\
\begin{equation}\label{e.13}
\begin{array}{rcl}
  P_{11} (\omega, \omega') & = & \mathcal{L}_2 (\omega) \mathcal{L}_1 (\omega')
  \left[
  K_{11}+ K_{22} \alpha_1(\omega) \alpha_1(\omega') +
  K_{12}  \alpha_1 (\omega')
 +  K_{21}  \alpha_1 (\omega)
    \right],\\\\
 P_{12} (\omega, \omega') & = & \mathcal{L}_2 (\omega) \mathcal{L}_1 (\omega')
  \left[
  K_{11} \alpha_2(\omega') + K_{22} \alpha_1(\omega)  +
   K_{12} + K_{21} \alpha_1 (\omega)  \alpha_2 (\omega')
    \right],\\\\
 P_{21} (\omega, \omega') & = & \mathcal{L}_1 (\omega) \mathcal{L}_2 (\omega')
  \left[
  K_{22} \alpha_1(\omega') + K_{11} \alpha_2(\omega)  +
   K_{21} + K_{12} \alpha_2 (\omega)  \alpha_1 (\omega')
    \right],\\\\
  P_{22}(\omega, \omega') & = & \mathcal{L}_1 (\omega) \mathcal{L}_2 (\omega')
  \left[
  K_{22}+ K_{11} \alpha_2(\omega) \alpha_2(\omega') +
  K_{21}  \alpha_2 (\omega')
 +  K_{12}  \alpha_2 (\omega)
    \right],
\end{array}
\end{equation}
\\
being
\\
\begin{equation}\label{e.14}
\mathcal{L}_i (\omega) \equiv \frac{t_i(\omega)}{D(\omega)} ,
\qquad \alpha_i(\omega) \equiv r_i(\omega) e^{i \omega l /c},
\qquad i = 1,2.
\end{equation}
\\
Now we evaluate the ratio $\mathcal{R}_{\out}(R,L|R,R)$ between
the probability  $P_{\out}(R,L)$ of observing one photon behind
mirror 2 and one photon behind mirror 1 (coincidence), and the
probability  $P_{\out}(R,R)$ of observing two photon behind mirror
2.\\ Using Eqs. (\ref{e.10}) we obtain
\\
\begin{equation}\label{d.108}
\begin{split}
\mathcal{R}_{\out}(R,L|R,R) & =
\frac{P_{\out}(R,L)}{P_{\out}(R,R)} = \left|
 \frac{\langle F_1(\eta), F_2(\eta) ; \out | \psi \rangle}
 {\langle F_1^2(\eta); \out | \psi \rangle} \right|^2 \\\\
 & = \left|
\frac{ \dis{\int d \omega_1 \int d \omega_2 \, \eta_1^*(\omega_1)
\eta_2^* (\omega_2) P_{12}(\omega_1,  \omega_2)}  }{\dis{ 2^{-1/2}
\int d \omega_1 \int d \omega_2 \, \eta_1^*(\omega_1) \eta_1^*
(\omega_2) P_{11}(\omega_1, \omega_2) }}
  \right|^2.
\end{split}
\end{equation}
\\
We now illustrate the meaning of this formula. We start writing
explicitally the  value of the ratio $P_{12}/P_{11}$:

 { \normalsize
\begin{equation}\label{e.15}
\frac{P_{12}(\omega_1,  \omega_2)}{P_{11}(\omega_1,  \omega_2)} =
\frac{ \dis{
  K_{11} r_2(\omega_2) e^{i \omega_2 l /c}  + K_{22}  r_1(\omega_1) e^{i \omega_1 l /c}  +
   K_{12} + K_{21} r_1(\omega_1) e^{i \omega_1 l /c}
     r_2(\omega_2) e^{i \omega_2 l /c}
   }  }{\dis{
  K_{11}+ K_{22}  r_1(\omega_1) e^{i \omega_1 l /c}
  r_1(\omega_2) e^{i \omega_2 l /c} +
  K_{12}  r_1(\omega_2) e^{i \omega_2 l /c}
 +  K_{21}  r_1(\omega_1) e^{i \omega_1 l /c}
}}.
\end{equation} }
\\
This expression is only apparently complicated, but each term at
numerator and denominator is susceptible to a clear physical
interpretation. We assume that $K_{ij}(\omega_1,\omega_2)$ is
proportional to the probability amplitude that a pair of excited
molecules within the cavity, emit spontaneously one photon with
angular frequency $\omega_1$ on mode $i$, and one photon with
angular frequency $\omega_2$ on mode $j$, being the
proportionality factor the same for all coefficients
$K_{ij}(\omega_1,\omega_2)$. More precisely we assume that
$|K_{ij}(\omega_1,\omega_2)|^2 d \omega_1 d \omega_2$ is
proportional to the emission probability of one photon on mode $i$
with angular frequency between $\omega_1$ and  $\omega_1 +  d
\omega_1$, and one photon on mode $j$ with angular frequency
between $\omega_2$ and  $\omega_2 +  d \omega_2$. At this point it
is easy to see how each term which appears in the ratio
(\ref{e.15}) admits a clear physical interpretation.
\\
 \begin{picture}(450,420)(-52,-10)
 \thinlines
  \put(40,350){\makebox(0,0){$ K_{12} + K_{21} r_1(\omega_1) e^{i \omega_1 l /c}
     r_2(\omega_2) e^{i \omega_2 l /c}$}}
    \put(40,250){\makebox(0,0){$ K_{11} r_2(\omega_2) e^{i \omega_2 l /c}
  + K_{22}  r_1(\omega_1) e^{i \omega_1 l /c}  $}}
    \put(40,150){\makebox(0,0){$K_{11}+ K_{22}  r_1(\omega_1) e^{i \omega_1 l /c}
  r_1(\omega_2) e^{i \omega_2 l /c}$ } }
    \put(40,50){\makebox(0,0){$ K_{12}  r_1(\omega_2) e^{i \omega_2 l /c}
 +  K_{21}  r_1(\omega_1) e^{i \omega_1 l /c} $ } }
 \multiput(250,50)(0,100){4}{\makebox(0,0){$+$}}
  \multiput(190,60)(0,100){4}{\vector(1,0){40}}
  \put(190,30){\vector(1,0){40}}
  \put(190,140){\vector(1,0){40}}
  \put(190,230){\vector(-1,0){40}}
  \put(190,340){\vector(-1,0){40}}
  \put(310,70){\vector(1,0){40}}
  \put(310,150){\vector(1,0){40}}
  \put(310,240){\vector(-1,0){40}}
  \put(310,330){\vector(-1,0){40}}
  \put(310,40){\vector(1,0){40}}
  \put(310,130){\vector(1,0){40}}
  \put(310,270){\vector(1,0){40}}
  \put(310,350){\vector(1,0){40}}
  \put(310.5,355){\oval(40,10)[l]}
  \put(309.5,335){\oval(40,10)[r]}
  \put(189.5,235){\oval(40,10)[r]}
  \put(310.5,265){\oval(40,10)[l]}
  \put(310.5,155){\oval(40,10)[l]}
  \put(310.5,135){\oval(40,10)[l]}
  \put(310.5,65){\oval(40,10)[l]}
  \put(190.5,35){\oval(40,10)[l]}
 \thicklines
 \multiput(165,20)(0,100){4}{\framebox(5,60)}
  \multiput(210,20)(0,100){4}{\framebox(5,60)}
   \multiput(285,120)(0,100){3}{\framebox(5,60)}
    \multiput(330,120)(0,100){3}{\framebox(5,60)}
   \multiput(285,20)(0,100){4}{\framebox(5,60)}
    \multiput(330,20)(0,100){4}{\framebox(5,60)}
\multiput(190,40)(0,100){4}{\circle*{3}}
\multiput(190,60)(0,100){4}{\circle*{3}}
\multiput(310,140)(0,100){3}{\circle*{3}}
\multiput(310,160)(0,100){3}{\circle*{3}}
\multiput(310,40)(0,100){4}{\circle*{3}}
\multiput(310,60)(0,100){4}{\circle*{3}}
\end{picture}

FIG. 5. Diagrams illustrating the probability amplitudes (reported
in the left column), relative to Eq. (\ref{e.15}) Here
$r_1(\omega)$ [$r_2(\omega)$] is the reflection coefficient of
mirror 1 (at the left) [2 (at the right)]. The photon of angular
frequency $\omega_1$ is always plot higher then photon of angular
frequency $\omega_2$.
\\
\\
With the help of  Fig. 5 we can see, e.g., that the first term at
numerator, corresponding to the first diagrams of the second row,
give us the probability amplitude of simultaneous emission of a
pair of photons toward right, and that the photon of angular
frequency $\omega_1$ is detected behind the mirror $2$, while the
photon of angular frequency $\omega_2$ is detected behind the
mirror $1$ after reflection on the mirror $2$. The transmission
coefficients and all contributes generated by multiple reflections
on the cavity mirrors, are computed into terms  $ \mathcal{L}_2
(\omega_1) \mathcal{L}_1 (\omega_2)$, which we have simplified
into Eq. (\ref{e.15}). All the other terms in Eq. (\ref{e.15}),
admit analogue interpretation shown by remaining diagrams in Fig.
5.\\ Of course, for reasons of internal consistency of the theory,
it need renormalize the coefficients $K_{ij}(\omega_1,\omega_2)$
imposing
\\
\begin{equation}\label{nuova}
\sum_{i, j}^{1,2} \int d \omega_1  \int d \omega_2
|K_{ij}(\omega_1,\omega_2)|^2  = 1.
\end{equation}
\\
At this point it is easy to obtain the correct probability
distributions. If we define
\\
\begin{equation}\label{nuova2}
\begin{split}
\mathcal{R}_\out(R,L|R,R) & \equiv \mathcal{R}_1, \\\\
\mathcal{R}_\out(R,L|L,L) & \equiv \mathcal{R}_2,
\end{split}
\end{equation}
\\
and we impose the normalization condition
\\
\begin{equation}\label{nuova2.1}
P_{\out}(R, R) + P_{\out}(R, L)  +   P_{\out}(L, L) =  1,
\end{equation}
\\
the desired distributions are then obtained after some algebra in
the form
\\
\begin{equation}\label{nuova3}
\begin{split}
P_{\out}(R, R) & = \frac{
 \mathcal{R}_2}{\mathcal{R}_1
 + \mathcal{R}_2+ \mathcal{R}_1 \mathcal{R}_2}, \\\\
P_{\out}(R, L) & = \frac{\mathcal{R}_1
 \mathcal{R}_2}{\mathcal{R}_1
  + \mathcal{R}_2+ \mathcal{R}_1 \mathcal{R}_2}, \\\\
 P_{\out}(L, L) & = \frac{\mathcal{R}_1
}{\mathcal{R}_1
 + \mathcal{R}_2+ \mathcal{R}_1 \mathcal{R}_2},
\end{split}
\end{equation}
\\
where, for example, $P_{\out}(R, R)$ is the normalized probability
to find two photon outside the cavity behind the  mirror $2$ of
the cavity.
\subsection{Single mode}
Now we suppose that the field mode spectrum is discretized by an
appropriate procedure  \cite{blow90}, furthermore we fix the
attention on  a single mode of assigned angular frequency
 $\omega$. The commutation relations for creation and annihilation
 operators defined on this discrete set of modes, are written as
\\
\begin{equation}\label{nuovo4}
 [\opa_i, \opa_j^\dag  ] =
    [\opc_i, \opc_j^\dag  ] = \delta_{i j},
\end{equation}
\begin{equation}\label{nuovo5}
  [\opb_i, \opb_j^\dag  ] = G_{i j}.
\end{equation}
\\
Now we can have two photon on a single discrete mode, therefore we
define the states generated by bilinear and quadratic forms of
{\em Inside} and {\em Outside} operators, as
\\
\begin{equation}\label{nuovo6}
| f_i, f_j ; \ing \ra \equiv \left( 2^{-1/2} \right)^{\delta_{i
j}} \opb_i^\dag \opb_j^\dag | 0 \ra ,
\end{equation}
\\
and
\\
\begin{equation}\label{nuovo7}
\begin{split}
| f_i, f_j ; \out \ra & \equiv \left( 2^{-1/2} \right)^{\delta_{i
j}} \opc_i^\dag \opc_j^\dag | 0 \ra \\\\
 & = \left( 2^{-1/2} \right)^{\delta_{i
j}} \sum_{k, l}^{1,2} \left( 2^{-1/2} \right)^{-\delta_{k l}} M_{i
k}^* M_{j l}^* | f_k, f_l ; \ing \ra.
\end{split}
\end{equation}
\\
respectively. It is easy to see that
\\
\begin{equation}\label{nuovo8}
\la f_i, f_j ; \ing| f_k, f_l ; \ing \ra = \left( 2^{-1/2}
\right)^{\delta_{i j} + \delta_{k l}}\left(  G_{i k}  G_{ j l} +
G_{i l}  G_{ j k} \right) .
\end{equation}
\\
Exactly as before, we define the most general two-photon state
created by {\em Inside} operators as
\\
\begin{equation}\label{nuovo9}
| \psi \ra = \sum_{i, j}^{1,2}  K_{i j} | f_i, f_j ; \ing \ra ,
\end{equation}
\\
where $\mathcal{K} = \mathcal{K}^{\mathrm{T} }$, and we calculate
\\
\begin{equation}\label{nuovo10}
\la  f_a, f_b ; \out | \psi \ra = 2 \left( 2^{-1/2}
\right)^{\delta_{a b}}  \left( \mathcal{M} \cdot \mathcal{G} \cdot
\overline{\mathcal{K}} \cdot \mathcal{G}^{\mathrm{T}} \cdot
\mathcal{M}^{\mathrm{T}} \right)_{a b},
\end{equation}
\\
where we have defined the matrix $\overline{\mathcal{K}}$ as
\\
\begin{equation}\label{nuovo11}
[\overline{\mathcal{K}}]_{i j} \equiv \overline{K}_{i j} = \left(
2^{-1/2} \right)^{\delta_{i j}} K_{i j}.
\end{equation}
\\
Therefore Eq. (\ref{e.13}) is formally still valid if we make the
substitution $K_{i i}\rightarrow K_{i i}/ \sqrt{2}$.
\\
From this point we consider the case of a symmetrical cavity, that
is we assume $r_1(\omega)=r_2(\omega) \equiv r(\omega)$ and
$t_1(\omega)=t_2(\omega) \equiv t(\omega)$. For simplicity we
choose the phase of transmission and reflection coefficients, as
\\
\begin{equation}\label{d.91.2}
  t(\omega) = i \sqrt{1 - R} \qquad \text{and} \qquad r(\omega) = -
  \sqrt{R},
\end{equation}
\\
and redefine, into a more expressive form,
\\
\begin{equation}\label{nuovo11}
K_{11} \equiv C_{RR}, \qquad K_{22} \equiv C_{LL}, \qquad K_{12} +
K_{21}  = 2 K_{12} \equiv C_{RL}.
\end{equation}
\\
Then in discrete mode representation, the normalization
conditions, can be written as
\\
\begin{equation}\label{nuovo12}
|C_{RR}|^2 + |C_{RL}|^2 + |C_{LL}|^2 = 1.
\end{equation}
\\
Finally we can write:
\\
\begin{equation}\label{d.110}
\mathcal{R}_{\out}(R, L|R, R) =  \left| \frac{\sqrt{2}( C_{R
R}+C_{R R} ) r e^{i \omega l/c} + C_{R L}(1 + r^2 e^{2 i \omega
l/c} ) } { C_{R R}  +  C_{L L} r^2 e^{2 i \omega l/c} + \sqrt{2}
C_{R L} r e^{i \omega l/c}}
  \right|^2.
\end{equation}
\\
We note  the presence of factors $\sqrt{2}$, which we have
introduced because of different normalization required by discrete
mode spectrum. Because of them, there is no exact correspondence
between diagram of Fig. 5 and terms into  Eq. (\ref{d.110}). This
factors arise since we are not using an orthonormal  base and
therefore a mixing between normalization of states with a photon
on  mode and two photons on  mode, is produced. Indeed we will
see, in the next section, that working with single photon states
which admit only a single normalization factor, it is possible to
obtain a direct association between diagrams and formulas.\\ Now
we calculate Eq. (\ref{d.110}) for some particular value of the
state $| \psi \ra$. Let  $| \psi \ra$ coinciding with each of the
three states of the orthonormal base defined in Appendix. It can
readily be shown that
\\

a) $ | \psi \rangle = | n_\pm \rangle $
\\
\begin{equation}\label{d.111}
 C_{R L} = \pm \frac{1}{\sqrt{2}}, \qquad C_{R R}=C_{L L}=\frac{1}{2}
 \qquad
\Rightarrow \qquad \mathcal{R}_{\out}(R, L|R, R) = 2.
\end{equation}
\\

b) $ | \psi \rangle = | n_0 \rangle $
\\
\begin{equation}\label{d.111.1}
 C_{R L} = 0, \qquad C_{R R}= - C_{L L}= - \frac{1}{\sqrt{2}}
 \qquad \Rightarrow  \qquad \mathcal{R}_{\out}(R, L|R, R) = 0.
\end{equation}
\\
It is remarkable that for these states having high symmetry,
$\mathcal{R}_{\out}$  does not depend either on mirrors
reflectivity, nor  on the phase  $\omega l /c$. Following our
interpretative scheme, Eq.  (\ref{d.111}) shows that when the
emission probability of the pair of photons on the same way or on
the opposite way, is the same, that is $|C_{R L}|^2=|C_{R
R}|^2+|C_{L L}|^2=1/2$, the probability to observe a coincidence
is twice with respect to the probability of not observing. Instead
when the two photon are emitted along a common way, but with pair
emission probability amplitude toward right and left which differs
for a sign, Eq. (\ref{d.111.1}), the probability of observing a
coincidence is zero. But within our interpretative scheme, which
require none distinction between left and right for emission of
the pair of the photon in a symmetrical cavity, it is hard to
think that this state really exist. Therefore a state  $|n_0
\rangle$ having $C_{R R} \neq C_{L L}$ is difficult to accept.
\\
From this point we consider only the case  $C_{R R}= C_{L L}$ and
we define
\\
\begin{equation}\label{d.112}
  \frac{C_{R R}}{C_{R L}} \equiv \zeta = | \zeta| e^{i \arg \zeta}.
\end{equation}
\\
Then using Eq.  (\ref{d.112}), Eq. (\ref{d.110}) can be written as
\\
\begin{equation}\label{d.112.1}
\begin{split}
& \mathcal{R}_{\out}(R, L|R, R) = \\\\ & \frac{8 R^2 z^2 - 4
\sqrt{2 R} \left[ \cos(x+y) + R \cos(x-y)\right] z + 1 + 2 R \cos
2 x + R^2 }{\left( 1 + 2 R \cos 2 x + R^2 \right) z^2 - 2 \sqrt{2
R} \left[ \cos(x-y) + R \cos(x+y) \right] z + 2 R },
\end{split}
\end{equation}
\\
where we have used the following notation
 \\
\begin{equation}\label{d.112.2}
  x \equiv \omega l /c, \qquad y \equiv \arg \zeta, \qquad z
  \equiv | \zeta |.
\end{equation}
\\
This expression seems still rather complicated, but we can learn
something from it considering the limit cases $C_{R R}=0
\Leftrightarrow z = 0$ and $C_{R L}=0 \Leftrightarrow z = \infty$:
\\
\begin{equation}\label{d.112.2}
\begin{split}
 \lim _{z \rightarrow 0} \mathcal{R}_{\out}(R, L|R, R) &
 = \frac{1+2 R \cos 2 x + R^2}{2 R} \equiv \mathcal{R}_0 , \\\\
  \lim _{z \rightarrow \infty} \mathcal{R}_{\out}(R, L|R, R) &
 = \frac{8 R^2}{1 + 2 R \cos 2x + R^2} \equiv \mathcal{R}_\infty.
\end{split}
\end{equation}
\\
The first of Eqs. (\ref{d.112.2}) is shown in Fig. 6. From Fig. 6b
we can see that when $R \rightarrow 0$, $ \mathcal{R}_0
\rightarrow \infty$ that is, from Eq. (\ref{d.110}), $P_{\out}(R,
R) \rightarrow 0$. Indeed if in the absence of the cavity the two
of photon are emitted one toward right and one toward left, it is
impossible to detect two from the same side. When $R \gtrsim 0.8$,
$\mathcal{R}_0$ is practically independent from  $R$ while it
presents an oscillation of period $ \pi$ in $x$. Observing Fig. 6a
is evident that when $R \gtrsim 0.5$, near the resonance
 $x \simeq \pi$ we have  $\mathcal{R}_0 \simeq
2$, while near $x = (2 n +1)\pi /2$, $n$ integer, there is a
region for which $\mathcal{R}_0 < 1$. This loss of coincidence is
due to the fact that only the first and the last pairs of diagrams
in Fig. 5. contribute to  $\mathcal{R}_0$ but in the first, which
gives the probability amplitude of observing the coincidence, the
two
 amplitudes $C_{RL}$ and  $C_{RL}R e^{2 i \omega l
/c}$ interfere destructively for $x \simeq (2 n +1)\pi /2$ and  $R
\simeq 1$ causing zero total amplitude.
The second of Eqs.  (\ref{d.112.2}) is shown in  Fig. 7a for
 $0 < \mathcal{R}_0 < 1$; the plane part of the
graph corresponds to the bigger than $1$ values. First of all we
observe an obvious fact: for $R = 0$, we have $\mathcal{R}_\infty
= 0$, that is if the two photons are emitted both on the same way,
it is impossible to observe a coincidence in absence of a cavity
that mixes the directions. From the graph it is evident as when
$R$ increment, the reflections increment too, and the coincidence
probability arise from the zero value. In  Fig. 7b the behaviour
of $\mathcal{R}_\infty $ is shown for $\mathcal{R}_\infty \leq 50
$. We note that when $x = (2 n +1)\pi /2$ and $R \rightarrow 1$,
we have $\mathcal{R}_\infty \rightarrow \infty$, that is the
probability of observing a pair of photons from a side of the
cavity, goes to zero. Indeed the second and the third diagram of
Fig. 5 contribute to $\mathcal{R}_\infty $ but the third diagram,
giving the probability of observation two parallel photons, goes
to zero for $x = (2 n +1)\pi /2$ and $R \simeq 1$ because of
destructive interference between $C_{RR}$ and  $C_{LL}R e^{2 i
\omega l /c}$. Therefore for  $z \gg 1$ and realistic
reflectivity, the probability of observing coincidence is always
bigger then probability of observing two photon on the same side
of the cavity. \\ On the other hand, we have also seen that for $z
\rightarrow 0$, $\mathcal{R}_\out$ can be less then $1$ for $R
\rightarrow 1$
and this is certainly the most interesting case to
investigate.
In Fig. 8 we have shown the behaviour of Eq. (\ref{d.112.1}), as
function of   $x$ and $y$, for several values $z$ and $R=.999$.
For $R \gtrsim 0.8$ the dependence from $R$ is negligible. In Fig.
9 we shown the contour  plot of
 $\mathcal{R}_\out$, between $0$ and $1$. From this figure it is evident
that the dark zones, corresponding to  $\mathcal{R}_\out < 1$,
have an extension gradually decreasing for $z$ increasing, until
they disappear for $z \geq 1$ (not shown in Fig. 9).
It is interesting to note that the probability of observing two
photon on one side of the cavity is bigger than coincidence
probability, when within the cavity the two photons are emitted
along opposite way.
Indeed only third and fourth diagrams in Fig. 5 contribute to
$P_{\out}(R, R)$, but while in the fourth diagram the two
amplitudes are always in phase, in the third diagram the two
amplitude $C_{RR}$ and $C_{LL} R e^{2 i \omega l /c}$ can have
opposite phase and interfere destructively for $R \sim 1$.
Then if $ |C_{R R}| > |C_{R L}|$, that is if $z > 1$, either the
third or the fourth diagram, give negligible contributions and
$P_{\out}(R, R) \sim 0$. Instead if $ |C_{R L}| > |C_{R R}|$ ($z <
1$) the fourth diagram gives a consistent contribution to
$P_{\out}(R, R)$ and at the same time the first diagram
(proportional to $C_{R L}$ but in condition of destructive
interference) gives a negligible contribution to $P_{\out}(R, L)$.
\\
Another interesting case is that of the resonance in a broad
sense, that is $\omega l /c = \pi N$, with $N$ integer: for $N$
odd there is resonance in a strict sense  while for  $N$ even
there is anti-resonance. In this case Eq. (\ref{d.112.1}) can be
simplified and written as
\\
\begin{equation}\label{d.113}
\mathcal{R}_{\out}(R, L|R, R) = \frac{ 4 F^2 z^2 - (-1)^{N} 4 F z
\cos y + 1} {z^2 - (-1)^{N} 2 F z \cos y + F^2} ,
\end{equation}
\\
where $F \equiv \sqrt{2 R}/(1+R)$. In the  Fig. 10a a plot of Eq.
(\ref{d.113}) is shown as function of $z$ and $y$, for $R = .5$
and $N$ odd. The analogue plot for  $N$ even can be obtained
translating the plot by an amount $\pi$ along the $y$ axes. It is
evident that the ratio $\mathcal{R}_{\out}$ is always greater then
$1$ except for a small region centered around $y= \pi$ and $z = 1/
2 F$ which disappear for $R \rightarrow 1$. We can always  write
Eq. (\ref{d.113}) as a ratio between two second degree polynomial
in $z$
\\
\begin{equation}\label{d.114}
\mathcal{R}_{\out}(R, L|R, R) = 4 F^2 \frac{(z- z_+^u )(z -
z_-^u)}{(z- z_+^d )(z -z_-^d)} ,
\end{equation}
\\
where we have defined the  roots of the two  polynomials as
\\
\begin{equation}\label{d.115}
z_\pm^u \equiv -\frac{1}{2 F}e^{\pm i y}, \qquad z_\pm^d \equiv -F
e^{\pm i y}.
\end{equation}
\\
Only for $y = \pi \, ( \mathrm{mod}\, 2 \pi)$ we can have real
positive root:
\\
\begin{equation}\label{d.116}
\mathcal{R}_{\out}(R, L|R, R) \Big|_{ y = \pi} = \left( \frac{2 F
z -1}{z - F} \right)^2.
\end{equation}
In  Fig. 10b a plot of Eq. (\ref{d.116}) is shown for values of
 $z$ near to $1/2F$; from this plot we can see in detail the
 "jump" from the pole to the zero. From Fig. 10d we can observe
that for $R \rightarrow 1$
 the pole in $F$ and the zero in $1/2F$ tend to the common value
$1/\sqrt{2}$ compensating each other, so that $\mathcal{R}_{\out}
= 2$. The distance between the pole and the zero decreases as
$\sim (1-R)^2$ and already for $R=.9$ is less than a part on a
hundred. It is reasonable to think that for higher and more
realistic reflectivity, it is not possible to generate really a
state so well defined to discriminate between the pole and the
zero. Furthermore the really physical situation is always
described by a continuous  superposition of modes, therefore we
think that effective value of $\mathcal{R}_{\out}(R, L|R, R)
\Big|_{ y = \pi}$ is $\sim 2$ in all plane $y-z$. At last we note
that when $C_{R R}=0$ or $C_{R L}=0$, we have respectively
\\
\begin{equation}\label{d.117}
 \lim _{z \rightarrow 0} \mathcal{R}_{\out}(R, L|R, R)
 = \frac{1}{F^2} \xrightarrow{R \rightarrow 1} 2,
\end{equation}
\\
\begin{equation}\label{d.117.1}
  \lim _{z \rightarrow \infty} \mathcal{R}_{\out}(R, L|R, R)
 = 4 F^2 \xrightarrow{R \rightarrow 1} 2.
\end{equation}
\\
From  Eqs. (\ref{d.117}-\ref{d.117.1}) we deduce that in these
conditions is physically indifferent if the two photons are
emitted in the same or in the opposite way within the cavity. We
have already obtained the  result  $\mathcal{R}_{\out}=2$, when
$|\psi \rangle = | n_\pm \rangle$ independently from $R$,
corresponding to equal probability of emission of a pair of photon
along the same way or in opposite way. In the present case we have
the same result for $\omega l /c = \pi N$ and $ R \sim 1$. This is
consistent with the fact that in the limit of total reflectivity
in which all frequencies satisfy the resonance (in a broad sense)
conditions $\omega_n = n \pi c/l$, $n$ integer,  it is impossible
to speak of direction (left or right) of emission of a photon,
because of the two counter-propagating wave  that constitute a
stationary wave within the cavity have exactly the same weight.
Finally we note that since $P_{\out}(R, R) = P_{\out}(L, L) $,
from Eqs (\ref{nuova2.1}-\ref{nuova3}) follows
\\
\begin{equation}\label{d.119}
P_{\out}(R, L) = \frac{\mathcal{R}_{\out}}{2 +
\mathcal{R}_{\out}}, \qquad P_{\out}(R, R) = \frac{1}{2 +
\mathcal{R}_{\out}}.
\end{equation}
\\
Then when $\mathcal{R}_{\out} = 2$ we have
\\
\begin{equation}\label{d.120}
P_{\out}(R, L) = \frac{1}{2 }, \qquad P_{\out}(R, R) =
\frac{1}{4},
\end{equation}
\\
in qualitative agreement with Ref. \cite{deangelis99}.
\subsection{Single photon states}
In this section, we started with investigation of two-photon
states, because we were interested at the leak of symmetry in the
photon-number probability distributions. Nevertheless the study of
one-photon states is not void of interest. Indeed in previous
subsection we have shown that in discrete mode representation, the
interpretation of the results, were "contaminated" from factors
$\sqrt{2}$ generated by  mixing between normalization of states
with one photon for mode (e.g. $|1, 1 \ra$) and two photons for
mode (e.g. $|2, 0 \ra$). Now we will see that working with
one-photon states this mixing never appear. In the most general
form, the one-photon state $| \phi \ra$ generated by {\em Inside}
operators, can be written as
\\
\begin{equation}\label{unf.1}
| \phi \ra = \sum_{i =1}^2 \int d \omega \, K_i (\omega) |
F_i(\omega); \ing \ra,
\end{equation}
\\
while the analogous state generated by {\em Outside} operators, is
given by
\\
\begin{equation}\label{unf.2}
\begin{split}
| F_a (\eta); \out \ra & = \opC_a^\dag(\eta) | 0 \ra \\\\ & = \int
d \omega \, \eta_a(\omega ) | F_a (\omega); \out \ra.
\end{split}
\end{equation}
\\
The  probability amplitude to find the electromagnetic field,
represented by the state $|\phi\ra$ within the cavity, in the
 state $| F_a (\eta) ; \out \ra $,
 is
\\
\begin{equation}\label{unf.3}
\la F_a (\eta); \out | \phi \ra = \sum_{i =1}^2 \int d \omega \,
\eta_a^* (\omega) \left[ \mathcal{M}(\omega) \mathcal{G}(\omega)
\right]_{a i} K_i(\omega),
\end{equation}
\\
The ratio between the probability $P_{\out}(R)$ of observing a
photon behind mirror $2$ and the probability $P_{\out}(L)$ of
observing a photon behind mirror $1$ is equal to
\\
\begin{equation}\label{unf.4}
\begin{split}
\mathcal{R}_{\out}(R|L) & = \frac{P_{\out}(R)}{P_{\out}(L)} =
\left|
 \frac{\langle F_1(\eta) ; \out | \phi \rangle}
 {\langle F_2(\eta); \out | \phi \rangle} \right|^2 \\\\
 & = \left|
\frac{ \dis{\int d \omega \, \eta_1^*(\omega)
\mathcal{L}_2(\omega) \left[ C_R(\omega) + C_L(\omega) r_1(\omega)
e^{i \omega l /c} \right]} }{\dis{  \int d \omega \,
\eta_2^*(\omega) \mathcal{L}_1(\omega) \left[ C_L(\omega) +
C_R(\omega) r_2(\omega) e^{i \omega l /c} \right]
 }}
  \right|^2,
\end{split}
\end{equation}
\\
where we have redefined  $K_1(\omega) \equiv C_R(\omega)$ and
$K_2(\omega) \equiv C_L(\omega)$. Exactly as in section 5.2, if we
assume $ C_R(\omega)$ ($ C_L(\omega)$) proportional to the
probability amplitude that an active medium within the cavity emit
a photon of angular frequency $\omega$ toward right (left), each
terms into Eq. (\ref{unf.4}) admit a clear physical interpretation
illustrated in Fig. 11.
\\
Since in discrete mode representation $| f_i ; \ing \ra \equiv
\opb_i^\dag |0 \ra$, it is evident that passing to discrete case
 each term between square bracket in Eq. (\ref{unf.4}), remain
formally unchanged, without any $\sqrt{2}$ factor.\\ For sake of
consistency now we must impose
\\
\begin{equation}\label{unf.5}
  \int d \omega \, \left\{ |C_{R}(\omega)|^2  + |C_{L}(\omega)|^2
  \right\}= 1.
\end{equation}
\\
\begin{picture}(450,250)(-52,-30)
 \thinlines
    \put(40,150){\makebox(0,0){$ C_{R}(\omega) + C_{L}(\omega)
     r_1(\omega) e^{ i \omega l /c}$ } }
    \put(40,50){\makebox(0,0){$  C_{R}(\omega)
     r_2(\omega) e^{ i \omega l /c} + C_{L}(\omega) $ } }
 \multiput(250,50)(0,100){2}{\makebox(0,0){$+$}}
  \multiput(190,150)(120,-10){2}{\vector(1,0){40}}
  \multiput(190,50)(120,-10){2}{\vector(-1,0){40}}
  \put(309.5,45){\oval(40,10)[r]}
  \put(310.5,145){\oval(40,10)[l]}
%
 \thicklines
 \multiput(165,20)(0,100){2}{\framebox(5,60)}
  \multiput(210,20)(0,100){2}{\framebox(5,60)}
   \multiput(285,20)(0,100){2}{\framebox(5,60)}
    \multiput(330,20)(0,100){2}{\framebox(5,60)}
\multiput(190,50)(0,100){2}{\circle*{3}}
\multiput(310,50)(0,100){2}{\circle*{3}}
\end{picture}

FIG. 11.  Diagrams illustrating the probability amplitudes
(reported in the left column), relative to Eq. (\ref{unf.4}) Here
$r_1(\omega)$ [$r_2(\omega)$] is the reflection coefficient of
mirror 1 (at the left) [2 (at the right)].
\\
\\
In this case the normalized probability $P_{\out}(R)$ and
$P_{\out}(L)$ are given by
\\
\begin{equation}\label{unf.6}
P_{\out}(R) = \frac{\mathcal{R}_{\out}(R|L)}{1 +
\mathcal{R}_{\out}(R|L)}, \qquad P_{\out}(L) = \frac{1}{1 +
\mathcal{R}_{\out}(R|L)} .
\end{equation}
\section{Conclusion}
We have derived some simple relations for electromagnetic field
inside and outside an optical cavity, using a non-unitary
transformation between {\em Inside} and {\em Outside} operators.
The convenience of this approach lies in the fact that we do not
need to know any details of internal processes that generate the
two photon, for calculate the photon-number probability
distribution outside the cavity. Conversely we can obtain
information on internal processes, by comparing the calculate and
measured probability distribution. The method is a natural
extension to non-unitary transformation, of the usual method
employed, e.g., in the quantum theory of the lossless
Beam-Splitter. \cite{campos89}.
\newpage
\begin{center}
{ \bf ACKNOWLEDGMENTS}
\end{center}
I am grateful to Daniele Fargion for helpful discussion and
encouragement, and to Elena Cianci and to Fabio Palmieri for their
help in writing the manuscript. A special thank to Giovanni Di
Giuseppe for reading the manuscript.
\section*{Appendix}
In the section 5.2 we  introduced the three orthonormal states $ |
n_\pm \ra$ e $| n_0 \ra$ without derivation. This will be done in
this Appendix. \\ We rewritten the commutation relation for
operators $\opb_1(\omega)$ and $\opb_2(\omega)$ in discrete mode
representation and symmetrical cavity, as
\\
\begin{equation}\label{f.1}
\begin{array}{ccccl}
\left[ \opb_2(\omega),\opb_2^\dag(\omega) \right]& = &\left[
\opb_1(\omega),\opb_1^\dag(\omega) \right]& \equiv &\Delta(\omega)
,\\\\ \left[ \opb_2(\omega),\opb_1^\dag(\omega) \right]& = &\left[
\opb_1(\omega),\opb_2^\dag(\omega) \right]^*& \equiv &\rho
(\omega) \Delta(\omega),
\end{array}
\end{equation}
\\
where we have defined
\\
\begin{equation}\label{f.2}
\begin{split}
\Delta(\omega)  \equiv  & \frac{1 - R^2}{1 -2R \cos (2 \omega l
/c)+R^2},\\\\ \rho(\omega) \equiv & -2 \frac{\sqrt{R}}{1 +R}\cos (
\omega l /c).
\end{split}
\end{equation}
\\
Using a slightly different notation with respect to section 5.2,
we define
\\
\begin{equation}\label{d.94}
| \er, \el; \ing \rangle \equiv \frac{(\opb_1^\dag)^\er}{\sqrt{\er
!}} \frac{(\opb_2^\dag)^\el}{\sqrt{\el !}}  |0\rangle , \qquad \er
+ \el = 2,
\end{equation}
\\
where again the factor $({\mathcal{R} ! \mathcal{L}!})^{-1/2}$ is
due to the possible presence of two photon on a single mode. Of
course, since $\opb_1$ and $\opb_2^\dag$ does not commute, this
kets  not form an orthonormal base, but they are however linearly
independent. Indeed if we define
\\
\begin{equation}\label{d.95}
  \langle \er, \el ; \ing | \er', \el'; \ing  \rangle
   \equiv \tilde{G}(\er, \el; \er', \el'),
\end{equation}
\\
we can calculate, using Eqs. (\ref{f.1}-\ref{f.2}),
\\
\begin{equation}\label{d.96}
 \tilde{G}(\er, \el; \er', \el') = \Delta^2
  \begin{pmatrix}
    1 & \sqrt{2}\rho & \rho^2 \\\\
   \sqrt{2}\rho & 1+\rho^2 & \sqrt{2}\rho \\\\
    \rho^2 & \sqrt{2}\rho & 1
  \end{pmatrix}.
\end{equation}
\\
Therefore the kets defined in Eq. (\ref{d.94}) are linearly
independent being their Gram determinant  positive \cite{smv3.1}:
 \\
\begin{equation}\label{d.98}
 \text{Det} [\tilde{G}] = \Delta^6 (1 - \rho^2)^3 \geq 0.
\end{equation}
\\
By diagonalization of  $\tilde{G}$, after some algebra we obtain
the orthonormal base we look for:
\\
\begin{equation}\label{d.99}
 \begin{split}
| n_+ \rangle & = \frac{1}{2} | 2, 0 ;\ing \rangle +
\frac{1}{\sqrt{2}} | 1, 1; \ing \rangle + \frac{1}{2} | 0, 2; \ing
\rangle , \\\\
 | n_0 \rangle & = - \frac{1}{\sqrt{2}} | 2, 0 ;\ing \rangle +
 \frac{1}{\sqrt{2}} | 0,2; \ing
\rangle , \\\\
 | n_- \rangle & = \frac{1}{2} | 2, 0 ;\ing \rangle -
\frac{1}{\sqrt{2}} | 1, 1; \ing \rangle + \frac{1}{2} |0,2; \ing
\rangle .
 \end{split}
\end{equation}
\\
\subsection*{Note on matrix $\tilde{G}$}
The form of the matrix $\tilde{G}$ is particular and justify this
little note. Let $\bS_0$, $\bS_1$ and $\bS_2$ the following $3
\times 3$ matrix:
\\
\begin{equation}\label{n.1}
\bS_0 =\begin{pmatrix}
  1 & 0 & 0 \\
  0& 1 & 0 \\
  0 & 0 & 1
\end{pmatrix}, \quad
\bS_1 =\begin{pmatrix}
  0 & 0 & 1 \\
  0& 1 & 0 \\
 1 & 0 & 0
\end{pmatrix}, \quad
\bS_2 =\begin{pmatrix}
  0 & 1 & 0 \\
  1& 0 & 1 \\
  0 & 1 & 0
\end{pmatrix}.
\end{equation}
\\
It can be readily shown that they satisfy the following
multiplication table:
\\
\begin{equation}\label{n.2}
\begin{array}{c|ccc}
 \bS_\mu \cdot \bS_\nu  & \bS_0 & \bS_1 & \bS_2 \\ \hline
 \bS_0 & \bS_0 & \bS_1 & \bS_2 \\
 \bS_1 & \bS_1 & \bS_0 & \bS_2 \\
 \bS_2 & \bS_2 & \bS_2 & \bS_0 + \bS_1 \\
\end{array}
\end{equation} 
\\
Now consider the generic matrix $\mathcal{N}(\alpha)$, $(\alpha
\in \mathbb{R})$ given by
\\
\begin{equation}\label{n.3}
\mathcal{N}(\alpha) = \bS_0 + \alpha^2 \bS_1 + \sqrt{2} \alpha
\bS_2.
\end{equation} 
\\
It is characterized by
\\
\begin{equation}\label{n.3.1}
\text{Det}[\mathcal{N}(\alpha)] = (1-\alpha^2)^3, \qquad
\text{Tr}[\mathcal{N}(\alpha)] = 3 + \alpha^2.
\end{equation}
\\
If we indicate with $\lambda_0, \lambda_\pm$ his eigenvalues and
with $\bn_0, \bn_\pm$ the corresponding eigenvectors, we can write
\\
\begin{equation}\label{n.3.2}
  \begin{cases}
    \bn_+ = \frac{1}{2}(1,  \sqrt{2}, 1) &  \lambda_+ = (1 + \alpha)^2,
    \\ \\
    \bn_0 = \frac{1}{\sqrt{2}}(-1, 0, 1) &  \lambda_0 = (1 -
    \alpha^2), \\ \\
    \bn_- = \frac{1}{2}(1, -\sqrt{2}, 1) &  \lambda_- = (1 -   \alpha)^2.
  \end{cases}
\end{equation}
\\
Using Eq. (\ref{n.2}) it is easy to see that
\\
\begin{equation}\label{n.4}
\mathcal{N}(\alpha) \cdot \mathcal{N}(\beta) = (1+\alpha \beta)^2
\mathcal{N}\left( \frac{\alpha + \beta}{1 + \alpha \beta} \right).
\end{equation}
\\
Since $\mathcal{N}\left( 0 \right) = \bS_0$ is the identity
matrix, it is clear that the inverse of $\mathcal{N}(\alpha)$ is
still a matrix of the form (\ref{n.3}). Indeed putting $\beta =
-\alpha$ into Eq. (\ref{n.4}) we obtain
\\
\begin{equation}\label{n.5}
\mathcal{N}(\alpha) \cdot \mathcal{N}(- \alpha) =  (1-\alpha^2 )^2
\bS_0,
\end{equation}
\\
that is
\\
\begin{equation}\label{n.6}
\mathcal{N}^{-1} (\alpha) = \frac{1}{(1-\alpha^2 )^2}
\mathcal{N}(- \alpha).
\end{equation}
\\
Finally, from Eq. (\ref{d.96}) we get
\\
\begin{equation}\label{n.7}
\tilde{G} = \Delta^2 \mathcal{N}(\rho) .
\end{equation}
\\
As a curiosity, we note that in two dimension the matrices
$\mathcal{N}(\alpha)$ which satisfy the algebra (\ref{n.4}) are
given by
\\
\begin{equation}\label{e.12}
\mathcal{N}(\alpha) \equiv \mathbf{\sigma}_0 + \alpha
\mathbf{\sigma}_1,
\end{equation}
\\
where $\mathbf{\sigma}_0 = \mathbb{I}$ and $\mathbf{\sigma}_1$ is
the first of the  Pauli matrices \cite[p. 160]{sakurai}. It is
easy to show that
\\
\begin{equation}\label{n.3.1}
\text{Det}[\mathcal{N}(\alpha)] = 1-\alpha^2, \qquad
\text{Tr}[\mathcal{N}(\alpha)] = 2.
\end{equation}
\\
The eigenvalues $ \lambda_\pm$ and the corresponding eigenvectors
$ \bn_\pm$ are given by
\\
\begin{equation}\label{n.3.2}
  \begin{cases}
    \bn_+ = \frac{1}{2}(1,  1) &  \lambda_+ = 1 + \alpha,
    \\ \\
    \bn_- = \frac{1}{2}(-1,  1) &  \lambda_- = 1 -   \alpha.
  \end{cases}
\end{equation}

\newpage

\newpage

\begin{center}
{\bf FIGURE CAPTIONS}
\end{center}

FIG. 6. Plots of $\mathcal{R}_0 $ for different range of values.
In (a) the plane part correspond to values of $\mathcal{R}_0$
greater than $2.5$.

\bigskip
\bigskip

FIG. 7. Plots of $\mathcal{R}_\infty $ for different range of
values. In (a) the plane part correspond to values of
$\mathcal{R}_\infty$ greater than $1$.

\bigskip
\bigskip

FIG. 8. Four plot of $\mathcal{R}_{\out} $ calculated for a
symmetrical cavity, $R=0.999$ and several values of $z$. The
dependence from  $R$ for $R \gtrsim 0.8$ is negligible and not
reported in the figure.

\bigskip
\bigskip

FIG. 9. Contour plot of corresponding plot in Fig. 8, shows for
values $\mathcal{R}_{\out} $ between $0$ and $1$. It is evident
that for  $z$ increasing, the zone on the plane $x-y$ in which
$\mathcal{R}_{\out} < 1 $ decreasing.

\bigskip
\bigskip

FIG. 10. (a) Plot of Eq. (\ref{d.113}) for $R = 0.5$. This is not
a  realistic value, but we choose it for show in a clear manner
the various quantities.
  (b) The same that in (a) for $y = \pi$; the vertical
straight lines passing through the pole in
   $F$ and through the zero in $1/2F$.
  (c) The same that in (a) for $z=1/2F$. (d) Plot of $1/2F$ (up)
   and  $F$ (down), as function of the reflectivity $R$.

\end{document}